\documentclass{article}

\usepackage[eandd, final]{neurips_2026}

\usepackage{hyperref}
\usepackage[utf8]{inputenc} 
\usepackage[T1]{fontenc}    
\usepackage{hyperref}       
\usepackage{url}            
\usepackage{booktabs}       
\usepackage{amsfonts}       
\usepackage{nicefrac}       
\usepackage{microtype}      
\usepackage{xcolor}         
\usepackage{subcaption}
\usepackage{graphicx}
\usepackage{booktabs}
\usepackage{multirow}
\usepackage{tabularx}
\usepackage{todonotes}
\usepackage{ragged2e}
\usepackage{algorithm}
\usepackage{algpseudocode}
\usepackage{float}
\usepackage{placeins}
\usepackage{listings}
\usepackage{tikz}
\usepackage{amsmath,amssymb,amsthm}
\usepackage{tikz}
\usepackage{graphicx}
\usepackage{amsmath,amssymb}
\usepackage{fontawesome5}
\usepackage{hyperref}

\usetikzlibrary{
    calc,
    arrows.meta
}

\usepackage[markup=underlined]{changes}

\definechangesauthor[name={Eugenie}, color=blue]{eugenie}
\setdeletedmarkup{\textcolor{red}{\sout{#1}}}

\setaddedmarkup{\textcolor{blue}{#1}}
\usepackage[inline]{enumitem}

\newcommand{\dd}{\mathrm d} 
\newcommand{\E}{\mathbb{E}}

\newtheorem{remark}{Remark}
\makeatletter
\@ifundefined{ack}{\newenvironment{ack}{\section*{Acknowledgments}}{}}{}
\makeatother
\title{DynamAuction: A Reinforcement Learning Environment for Repeated Auctions with Dynamic Value}

\author{%
  Eugénie Patard\textsuperscript{\normalfont 1,}\thanks{Equal contribution}
  \quad
  Corentin Pla\textsuperscript{\normalfont 1,2,3,}\footnotemark[1]
  \quad
  Benjamin Heymann\textsuperscript{\normalfont 1,2,}\thanks{Equal supervision}
  \quad
  Patrick Loiseau\textsuperscript{\normalfont 1,}\footnotemark[2]
  \\[0.5em]
  \textsuperscript{\normalfont 1}INRIA, FairPlay Joint Team  \\
  \textsuperscript{\normalfont 2} Criteo AI Lab \\
  \textsuperscript{\normalfont 3} CREST-ENSAE \\ Paris, France 
}
\begin{document}
\maketitle
\addtocontents{toc}{\protect\setcounter{tocdepth}{-1}}
\begin{center}
    \href{https://github.com/eugeniepatard/dynamauction}{
        \faGithub\ \texttt{DynamAuction}
    }
\end{center}
\begin{abstract}
A buyer who repeatedly participates in auctions for related items often faces a dynamic value problem: what they win today changes what tomorrow’s items are worth. Despite the empirical importance of such effects, especially in display advertising, the algorithmic literature on repeated auctions has largely focused on the learning problem, by treating each opportunity as a contextual bandit, while sidestepping the planning problem induced by the dependence of current value on past wins.
We introduce DynamAuction, a synthetic environment for single-buyer sequential bidding. The reward from each win depends on the buyer’s history of previous wins, summarized through a carry-over state.
The environment supports first- and second-price auctions, non-homogeneous Poisson arrivals,
delayed feedback, saturation, and recency effects.
We also introduce a new class of Model Predictive Control (MPC) algorithms. On our benchmark scenario, these MPC policies outperform off-the-shelf deep RL baselines (PPO, SAC, TD3) as well as a tuned shading baseline.
We release DynamAuction to support RL research on repeated auctions.
\end{abstract}
\section{Introduction}

\paragraph{Motivation.}
In many real-world bidding settings, a buyer participates in a long sequence of auctions for related items, and the marginal value of winning the next one depends on what they have already won. The simplest instance is a buyer with single-unit demand: a firm that needs to acquire exactly one piece of equipment over a quarter through a series of auctions derives no additional value from a second win, so the value of the next opportunity drops to zero as soon as one auction has been won.  More complex dynamics arise in display advertising. For example, showing the same banner to a user multiple times within a short period typically leads to diminishing returns, so the value of the next impression depends on the timing and outcomes of previous auctions.

\paragraph{Learning vs.\ planning in repeated auctions.} Most existing work on repeated auctions focuses on learning how much to bid from contextual information and an unknown competition distribution~\cite{weed2016online,achddou2021efficient,NIPS2017_0bed45bd}. The only planning component typically included is a global budget or ROI constraint, which couples bids over time through a Lagrangian or dual variable~\cite{NIPS2017_0bed45bd,10.5555/3618408.3619926,10.1145/3033274.3084088,balseiro2020dual}. 
Beyond this constraint, these approaches generally assume that the value of each opportunity depends only on its current context, without accounting for how past wins affect future values. In many applications, improving planning by accounting for these dynamics may be more beneficial than further improving learning under a static value model. Our work aims to study this question by introducing a reinforcement learning environment in which these dependencies can be explicitly modeled and evaluated.

\paragraph{Motivations for a synthetic environment.}
Evaluating bidding algorithms on real-world data presents several challenges. Logged auction data only reflect the decisions made by the bidding policy used to collect them, making it difficult to evaluate alternative strategies without additional assumptions. Online evaluation through A/B tests is the textbook alternative, but it is slow, costly, constrained by business and ethical risk, and does not let the researcher independently vary delay, saturation, recency, or interference. A simulator avoids these limitations by allowing each factor to be varied under controlled conditions. Although \textsc{DynamAuction} can be used with general reinforcement learning algorithms, our experiments focus on planning and on the value of accounting for past wins.

\paragraph{Contributions.}
\begin{itemize}
  \item We formalize a model of single-buyer sequential bidding in which the per-display reward is generated by a kernel conditioned on a \emph{carry-over context} obtained by aggregating past wins through a \emph{commutative monoid}. This framework accommodates delayed feedback, saturation and recency effects, exogenous interference, and user heterogeneity.
  \item We release \textsc{DynamAuction}, an open-source, Gymnasium-compatible implementation of this model, together with a default benchmark task and guidelines for users to define new tasks.
  \item We implement and evaluate a panel of baselines (constant bid, prediction-based bidding, greedy myopic bidding with shading, PPO, SAC, TD3) and introduce a class of Model Predictive Control algorithms. We show empirically that, on the benchmark scenario, this myopic-optimal MPC outperforms the deep-RL baselines and a tuned shading baseline.
\end{itemize}

\paragraph{Outline.}
Section~\ref{sec:related} reviews the related literature and motivates the design of the environment. Section~\ref{sec:model} presents the model, and Section~\ref{sec:env} introduces \textsc{DynamAuction}. Section~\ref{sec:mpc} describes our MPC approach. Section~\ref{sec:experiments} presents the experiments, and Section~\ref{sec:conclusion} concludes. We use display advertising as the main running example throughout the paper. Accordingly, we refer to a won auction as a \textit{display} and sometimes to its reward as a \textit{conversion}. 

\section{Related work}
\label{sec:related}
\subsection{Repeated auctions, attribution, and budget/ROI constraints}

Much of the literature on repeated auctions studies a single bidder facing an unknown competition distribution, using bandit~\cite{lattimore2020bandit} or online learning methods. For example, \cite{weed2016online,achddou2021efficient} consider repeated second-price auctions where the bidder learns the value of an item only after winning it. They develop bidding strategies with theoretical guarantees in stochastic and adversarial settings. Other works extend this framework to include global budget or ROI constraints, often using knapsack formulations~\cite{10.1145/3033274.3084088,balseiro2020dual,castiglioni2022unifying,10.5555/3692070.3692294,NIPS2017_0bed45bd,10.1145/3543507.3583491,heymann2019cost}. This perspective has received considerable attention in the theoretical literature and has led to several important modeling contributions~\cite{10.1145/3038912.3052682,balseiro2015repeated,conitzer2022multiplicative}. A recent survey is provided in~\cite{10.1145/3699824.3699838}.

A second line of work uses reinforcement learning (RL) to model sequential bidding as a Markov Decision Process (MDP). \cite{10.1145/3018661.3018702} formulate real-time bidding as an MDP whose state encodes the campaign's remaining budget and lifetime together with the current bid request, and learn a value-function approximation for the resulting bidding policy. \cite{10.1145/3269206.3271748} propose a model-free RL approach for budget-constrained bidding, while \cite{Yang_2019} treat bid optimization as a multivariable control problem for a set of campaigns. A more general RL framework with budget and KPI constraints is introduced in \cite{10.1145/3447548.3467199}. More recent work has questioned whether the current state contains enough information for effective bidding decisions. This has motivated generative approaches that model entire bidding trajectories rather than relying only on the current state~\cite{guo2024generative,10.1145/3746252.3761497}. However, these approaches mainly capture dependencies between bidding decisions through campaign-level constraints, such as budgets or ROI targets. The value of an impression is generally determined by its current context, without explicitly modeling how previous wins affect the value of future opportunities.

A third line of work focuses on \emph{attributing} observed conversions back to the displays that caused them, and on bidding directly on incremental rather than observed outcomes. The general framing of computational advertising as a counterfactual learning problem goes back to \cite{bottou2013counterfactual}. On the bidding side, \cite{DiemertMeynet2017} show empirically that attribution-aware bidding outperforms last-touch baselines, and \cite{10.1145/3447548.3467280} and \cite{bompaire2024fixed} develop causal models for repeated user interactions, the latter introducing the myopic-optimal valuation that we will reuse. \cite{lewis2022incrementality} formalize the link between incrementality and attribution. \cite{heymann2024repeated} introduce the concept of repeated auctions with dynamic value and consider a setting simple enough to be solved with a differential equation, \cite{heymann2026learning} leverage the analysis to build a no-regret learning algorithm based on plugin estimators,
and \cite{badanidiyuru2022incrementality} study incrementality bidding via reinforcement learning under mixed and delayed rewards.
These works highlight the importance of accounting for past interactions, diminishing returns, and delayed outcomes. Our environment allows these effects to be modeled explicitly and studied under controlled conditions.

\subsection{RL benchmarks, simulators, and datasets for auctions}
A growing collection of open resources targets the bidding problem in online advertising. We list below those most relevant to our work.

\textbf{Datasets.}
These contributions mostly cast the auction as a supervised learning problem.
\cite{10.1145/2648584.2648590,zhang2014real} release the iPinYou dataset, the first publicly available large-scale RTB log, consisting of $\approx 35$\,GB of bid, impression, click, and conversion logs from a 2013 DSP competition. No environment for off-policy evaluation is provided. \cite{DiemertMeynet2017} release the Criteo Attribution Modeling for Bidding dataset, which contains 30 days of real advertising data with attribution labels. The dataset can be used to study attribution and incrementality, but does not provide an environment for simulating auction outcomes under different bidding strategies.

\textbf{Simulators.}
\cite{jeunen2023off} release \textsc{AuctionGym}: a Python simulator of first- and second-price advertising auctions designed for the off-policy evaluation of bandit-style learning-to-bid algorithms. It provides configurable competition; however, the per-impression value is a function of the current context only.

\textbf{Datasets paired with simulators.}
More recently, \cite{su2024auctionnet} release the simulated environment of the Alibaba/NeurIPS\,2024 auto-bidding competition: it consists of a large dataset and an RL-ready state/action/reward interface designed for cost-per-action-constrained bidding.
\cite{10.1145/3696410.3714657} release \textsc{BAT}, a benchmark built on the Avito platform, comprising 10 million first-price and 1 million Vickrey–Clarke–Groves (VCG) auction records collected over a month, together with a simulator and budget-pacing and cost-per-click (CPC) constrained baselines.

None of these resources is designed to exhibit the dependence of a single bidder's per-opportunity value on the history of past wins. \textsc{DynamAuction} is therefore complementary: \textsc{AuctionGym}, \textsc{BAT}, and the AuctionNet environment are natural sandboxes for studying learning-to-bid and budget/ROI/Cost Per Auction (CPA)-constrained pacing; the iPinYou and Criteo datasets are appropriate for Click-Through Rate (CTR)/Conversion Rate (CVR) and attribution modeling on real logs; and \textsc{DynamAuction} is intended as a sandbox for studying planning under the dynamics induced by past wins on the value of future opportunities.

\subsection{Model Predictive Control and model-based RL}

Model Predictive Control (MPC) \cite{CamachoBordons2013} uses a predictive model to optimize a sequence of actions over a finite horizon, executes only the first action, and replans at the next decision step. By evaluating candidate actions through model predictions rather than direct interactions with the environment, MPC is particularly appealing in settings where real-world experimentation is costly, provided that a sufficiently accurate model is available. MPC is an instance of a broader class of model-based approaches that use learned or known dynamics to guide decision-making. In model-based reinforcement learning, these approaches are commonly distinguished by how and when the model is used. In \emph{decision-time planning}, the model is used at each decision step to evaluate candidate action sequences and select an action. PETS~\citep{chua2018deepreinforcementlearninghandful} is a representative example, as are the MPC algorithms proposed in Section~\ref{sec:mpc}. In \emph{background planning}, the model is used during training to simulate trajectories and improve a policy or value function. Unlike decision-time planning, these methods learn a policy using model-generated trajectories, rather than using the model to optimize actions at each decision step. MBPO~\citep{janner2021trustmodelmodelbasedpolicy} and Dreamer~\citep{hafner2020dreamcontrollearningbehaviors} follow this approach. MuZero~\citep{Schrittwieser_2020} combines elements of both: it learns a compact latent dynamics model used for decision-time planning through Monte Carlo tree search, while training the model to support value and reward predictions rather than reconstructing the full environment.  Model-based approaches have also been explored in real-time bidding, primarily to address budget allocation, pacing, and ROI or CPA constraints~\citep{10.1145/3018661.3018702,10.5555/3545946.3598810}. In contrast, our setting focuses on a different source of temporal dependence: rather than campaign-level constraints, past auction wins influence the future marginal value of targeting the same user. The closest related work is repeated bidding with dynamic valuations~\citep{heymann2024repeated}. Building on this perspective, \textsc{DynamAuction} provides a configurable benchmark for studying such sequential bidding decisions. 

\subsection{Benchmark Design Requirements}
\label{sec:desiderata}
Based on the literature reviewed above, we identify three key properties for benchmarking sequential bidding with dynamic valuations. \textit{Delayed feedback.} The reward from winning an auction may be observed only after a stochastic delay, which can be long. \textit{Diminishing returns.} The marginal value of winning an auction may decrease as the bidder accumulates wins for the same user, reflecting saturation effects. \textit{Recency effects.} The marginal value of a win may also depend on the time elapsed since previous wins, rather than only on their number. Beyond these three properties, the environment should allow for heterogeneous users, exogenous events affecting user behavior, different opponent bid distributions and auction formats, and configurable auction arrival processes. We do not consider campaign-level budget or ROI constraints, which have been extensively studied and are separate from the user-level dynamics considered here. They could nevertheless be incorporated in future extensions.

\section{Model: sequential bidding with carry-over value}
\label{sec:model}

\subsection{Buyer, opportunities, and arrival process}

We consider a single buyer interacting over a finite, possibly random, time horizon $[0,T]$ where $T>0$, with a user $u$ drawn from a set $\mathcal{U}$ according to a probability measure $\rho$. A \emph{bidding opportunity} for $u$ is an event at which the buyer is offered the chance to bid in an auction for an item\footnote{In the context of RTB auctions, this is a placement for an online banner on a publisher website.} described by a feature vector $x \in \mathcal{X}$, where $\mathcal{X}$ is a measurable feature space. Bidding opportunities for a fixed user $u$ arrive at random times $0 \le \theta_1 < \theta_2 < \cdots$ generated by a non-homogeneous Poisson process on $[0,T]$ with intensity $\mu : [0,T] \to \mathbb{R}_{\ge 0}$, and item feature $X_j \in \mathcal{X}$ is drawn from a conditional distribution $\nu(\cdot \mid u, \theta_j)$. The buyer observes $X_j$, places a bid $b_j \in \mathbb{R}_{\ge 0}$, and competes against an opponent's bid $B_j \in \mathbb{R}_{\ge 0}$ drawn independently from a competition distribution $F$.\footnote{It is standard to aggregate all the competition into a single price-to-beat.} The auction is resolved by an allocation rule $\mathcal{W}(b,B) \in \{0,1\}$ (winning indicator) and a payment rule $p(b,B) \in \mathbb{R}_{\ge 0}$. For instance, the first-price rule corresponds to $\mathcal{W}(b,B) = \mathbf{1}\{b > B\},\; p(b,B) = b\,\mathcal{W}(b,B)$ and the second-price rule corresponds to $\mathcal{W}(b,B) = \mathbf{1}\{b > B\},\; p(b,B) = B\,\mathcal{W}(b,B)$. We let $W_j \in \{0,1\}$ denote whether opportunity $j$ is won. We write $t_1 < t_2 < \cdots$ for the corresponding \emph{win times}, that is, the opportunity times $\theta_j$ such that $W_j = 1$. Finally, $x_i$ denotes the feature of the item won at time $t_i$. User heterogeneity is represented through the dependence of the model primitives on $u$: the arrival intensity $\mu(\,\cdot\, | u)$, the item-feature distribution $\nu(\,\cdot \mid u, \theta)$, the law of the exogenous process $\Pi_u$, and the reward kernel $\mathcal{K}(u, \cdot)$ may all depend on $u$ (see below for the definition of these terms). The population measure $\rho$ on $\mathcal{U}$ then determines how this heterogeneity aggregates into the buyer's objective \eqref{eq:total-reward}. For $t \in [0,T]$, the set of win indices up to time $t$ is $\mathcal{I}_u(t) = \{i : t_i \le t\}$, and we write $\mathcal{I}_u(t^-) = \{i : t_i < t\}$ for the wins strictly before $t$.

\subsection{Carry-over context as a commutative monoid}
\label{subsec:carry-over}

To capture the dependence of the buyer's per-opportunity value on past wins, we summarize the user's display history into a single state. We fix a (measurable) space\footnote{In what follows, we omit most of the measurability assumptions.} $\mathcal{S}$ together with a commutative monoid\footnote{A monoid is a set equipped with an associative binary operation and an identity element.} structure $(\mathcal{S}, \oplus, \mathbf{0})$. The binary operation $\oplus : \mathcal{S} \times \mathcal{S} \to \mathcal{S}$ is associative and commutative, and $\mathbf{0} \in \mathcal{S}$ is its identity element. We further fix a \emph{decay map}
      $\gamma : \mathcal{X} \times [0,T] \times [0,T] \to \mathcal{S}, \quad (x, t', t) \mapsto \gamma(x, t', t)$,
intended to be evaluated only at $t' \le t$, and which encodes the residual contribution at time $t$ of a display with feature $x$ that was shown at the earlier time $t'$.
For a fixed user $u \in \mathcal{U}$ with realized win sequence $(t_i, x_i)_{i \in \mathcal{I}_u(t)}$ at time $t \in [0,T]$, we define the \emph{carry-over context} at time $t$ as
\begin{align}
\label{eq:carry-over}
  S_u(t) \;=\; \bigoplus_{i \in \mathcal{I}_u(t)} \gamma(x_i, t_i, t) \;\in\; \mathcal{S},
\end{align}
with the empty sum convention $S_u(t) = \mathbf{0}$ when $\mathcal{I}_u(t) = \emptyset$. Since $\oplus$ is associative and commutative, the value of the sum does not depend on the order in which past wins are aggregated. By construction, $S_u$ is c\`adl\`ag in $t$: it jumps at each win time $t_i$ and evolves smoothly between wins, through the dependence of $\gamma$ on its third argument. Familiar instances of $(\mathcal{S},\oplus,\mathbf{0})$ that we use
in the experiments include $(\mathbb{R}_{\ge0},+,0)$ for additive
counters, $([0,1],\times,1)$ for multiplicatively decreasing marginal
effects, and $([0,+\infty],\min,+\infty)$ for the
time since the most recent qualifying win. For the recency context, a qualifying display contributes
$\gamma(x,t',t)=t-t'$, while a nonqualifying display contributes
$+\infty$. Taking the minimum gives the time since the most recent qualifying win, or $+\infty$ if there has been none. A different choice is to retain the full display history.
We use $(\mathsf{H},\cup,\emptyset)$, where $\mathsf{H}$ is the
collection of finite sets of display--time pairs $(x_i,t_i)$.
Each display contributes $\gamma(x,t',t)=\{(x,t')\}$, and these
contributions are combined by set union. The timestamps allow us
to recover the chronological order of the displays.

\subsection{Reward kernel}
We define a \emph{reward kernel} that describes how the expected reward of a display is distributed over time. It is modeled as a positive measure over conversion delays, parametrized by the user, the displayed item, the carry-over context at the time of display, and an exogenous information process (introduced in Section~\ref{sec:exo}).
Formally, we define a measurable map
\[
  \mathcal{K} : \mathcal{U} \times \mathcal{X} \times \mathcal{S} \times \mathcal{Y} \times [0,T] \;\to\; \mathcal{M}_+\bigl([0,\infty)\bigr),
\]
where $\mathcal{Y}$ is the value space of the exogenous information process and $\mathcal{M}_+\bigl([0,\infty)\bigr)$ denotes the set of positive Borel measures on $[0,\infty)$. For a display occurring at time $t$, the kernel specifies the expected reward associated with different delays after the display. For any Borel measurable set $A \subseteq [0,\infty)$ of such delays, we write 
\begin{equation}
\label{eq:kernel-decomp}
  \mathcal{K}(u, x, s, y, t)(A) \;=\; \kappa(u, x, s, y, t)\delta_0(A) \;+\; \int_A \lambda(u, x, s, y, t; \tau)\,\dd\tau,
\end{equation}
where the Dirac component $\kappa$ models the \emph{instantaneous} part of the reward, while the absolutely continuous component $\lambda$ models the \emph{delayed} part as a density over the delay $\tau \ge 0$ since the display.  For a single won display at time $t_i$ on user $u$ with feature $x_i$, the expected reward attributable to this display and generated over a time window $[t_i, t_i + h]$ for $h \ge 0$ is
\begin{equation}
\label{eq:per-display-reward}
  r_i(h) \;=\; \kappa\bigl(u, x_i, S_u(t_i^-), y_u(t_i), t_i\bigr) \;+\; \int_0^{h} \lambda\bigl(u, x_i, S_u(t_i^-), y_u(t_i), t_i;\, \tau\bigr)\,\dd\tau,
\end{equation}
where $S_u(t_i^-)$ denotes the left-limit of the carry-over context at $t_i$, i.e., the carry-over context evaluated just \emph{before} the $i$-th win is incorporated into it.\footnote{The left-limit convention ensures that a winning bid is rewarded against the context that justified placing it, rather than against a context that already includes the consequence of the win itself.}
The buyer's expected cumulative reward over the horizon $[0,T]$ is then obtained by integrating per-display contributions over all wins occurring before $T$ against the population measure $\rho$,
\begin{equation}
\label{eq:total-reward}
  \mathcal{R}(T) \;=\; \int_{\mathcal{U}} \E\!\left[\,\sum_{i \in \mathcal{I}_u(T)} r_i(T - t_i)\,\right] \rho(\dd u),
\end{equation}
where the expectation is over the arrival process, the item-feature distribution, the competition, the exogenous process, and any randomness in the bidding policy, all conditional on the user $u$. The expected cumulative payment is
$
  \mathcal{C}(T) \;=\; \int_{\mathcal{U}} \mathbb{E}\!\left[\,\sum_{j : \theta_j \le T} p(b_j, B_j)\,\right] \rho(\dd u),
$
and the buyer's net cumulative profit $\mathcal{R}(T) - \mathcal{C}(T)$ is what they seek to maximize.

\subsection{Exogenous information and information available to the buyer}
\label{sec:exo}
\paragraph{Exogenous information process.}

For each user $u \in \mathcal{U}$, we allow exogenous events that affect the value of future displays but are independent of the buyer's actions. These events occur at times $0 \le \tau_1^u < \tau_2^u < \cdots$ and have marks $e_\ell^u \in \mathcal{E}$. We treat the marked point process $\Pi_u = (e_\ell^u, \tau_\ell^u)_{\ell \ge 1}$ as a primitive of the model and assume it is independent of the buyer's bidding policy. We then summarize the past of this process into a c\`adl\`ag \emph{exogenous information process}
\begin{align}
      y_u : [0,T] \to \mathcal{Y}, \qquad
  y_u(t) \;=\; \Psi\bigl(\{(e_\ell^u, \tau_\ell^u) : \tau_\ell^u \le t\}\bigr),
\end{align}
The resulting state $y_u(t)$ affects the reward generated by a display at time $t$ but evolves independently of the buyer's actions.

\begin{figure}[H]
\centering
\scalebox{0.8}{%
\begin{tikzpicture}[
    >=Stealth,
    every node/.style={font=\small\sffamily},
    arrow/.style={->, thick, blue!65!black}
]

\def\colgap{0.35cm}
\def\rowgap{0.35cm}

\pgfmathsetlengthmacro{\figwidth}{0.92\linewidth}
\pgfmathsetlengthmacro{\blockwidth}{(\figwidth-\colgap)/2}

\tikzset{
    block/.style={
        draw,
        rounded corners=2pt,
        minimum width=\blockwidth,
        text width=\dimexpr\blockwidth-14pt\relax,
        minimum height=2.2cm, 
        align=left,
        inner xsep=7pt,
        inner ysep=5pt,
        outer sep=0pt
    }
}

\node[
    draw,
    rounded corners=2pt,
    fill=gray!10,
    minimum width=\figwidth,
    inner xsep=7pt,
    inner ysep=5pt,
    outer sep=0pt
] (init) at (0,0) {%
    \shortstack[c]{%
        \textbf{Initialization}\\[3pt]
        Sample $u\sim\rho(\cdot)$, horizon $T$, auction times
        $\{\theta_j\}_{j\geq 1}\sim\mu(\cdot\mid u)$ \\[2pt]
       and  initial display $x_1\sim\nu(\cdot\mid u,\theta_1)$.
    }%
};

\node[
    block,
    fill=blue!7,
    anchor=north west
] (obs) at ([yshift=-\rowgap]init.south west) {%
    \textbf{1. Observation}\\[3pt]
    At opportunity $j$, the buyer observes
    \[ Z_j = \bigl( u,X_j,\phi(S_u(\theta_j^-),\theta_j),  y_u(\theta_j),\theta_j \bigr)  \]
};

\node[
    block,
    fill=green!8,
    anchor=north west
] (agent) at ([xshift=\colgap]obs.north east) {%
    \textbf{2. Buyer}\\[3pt]
    Select bid $b_j=\pi(Z_j)$.
};

\node[
    block,
    fill=violet!8,
    anchor=north west
] (update) at ([yshift=-\rowgap]obs.south west) {%
    \textbf{4. Environment update}\\[3pt]
    Update $S_u$ if $W_j=1$.\\
    Generate rewards via $\mathcal{K}$.\\
    Move to $\theta_{j+1}$.
};

\node[
    block,
    fill=orange!10,
    anchor=north west
] (auction) at ([xshift=\colgap]update.north east) {%
    \textbf{3. Auction}\\[3pt]
    Sample price to beat $B_j\sim F$; reveal the outcome:\\[2pt]
    $W_j=\mathcal{W}(b_j,B_j) \quad p_j=p(b_j,B_j)$.
};

\draw[arrow] (obs.east) -- (agent.west);
\draw[arrow] (agent.south) -- (auction.north);
\draw[arrow] (auction.west) -- (update.east);
\draw[arrow] (update.north) -- (obs.south);

\end{tikzpicture}%
}
\caption{
Overview of \textsc{DynamAuction}, illustrating the interaction between the bidding agent and the environment, the dependence of future values on past wins, and the timing of rewards.
}
\label{fig:env-schema}
\end{figure}
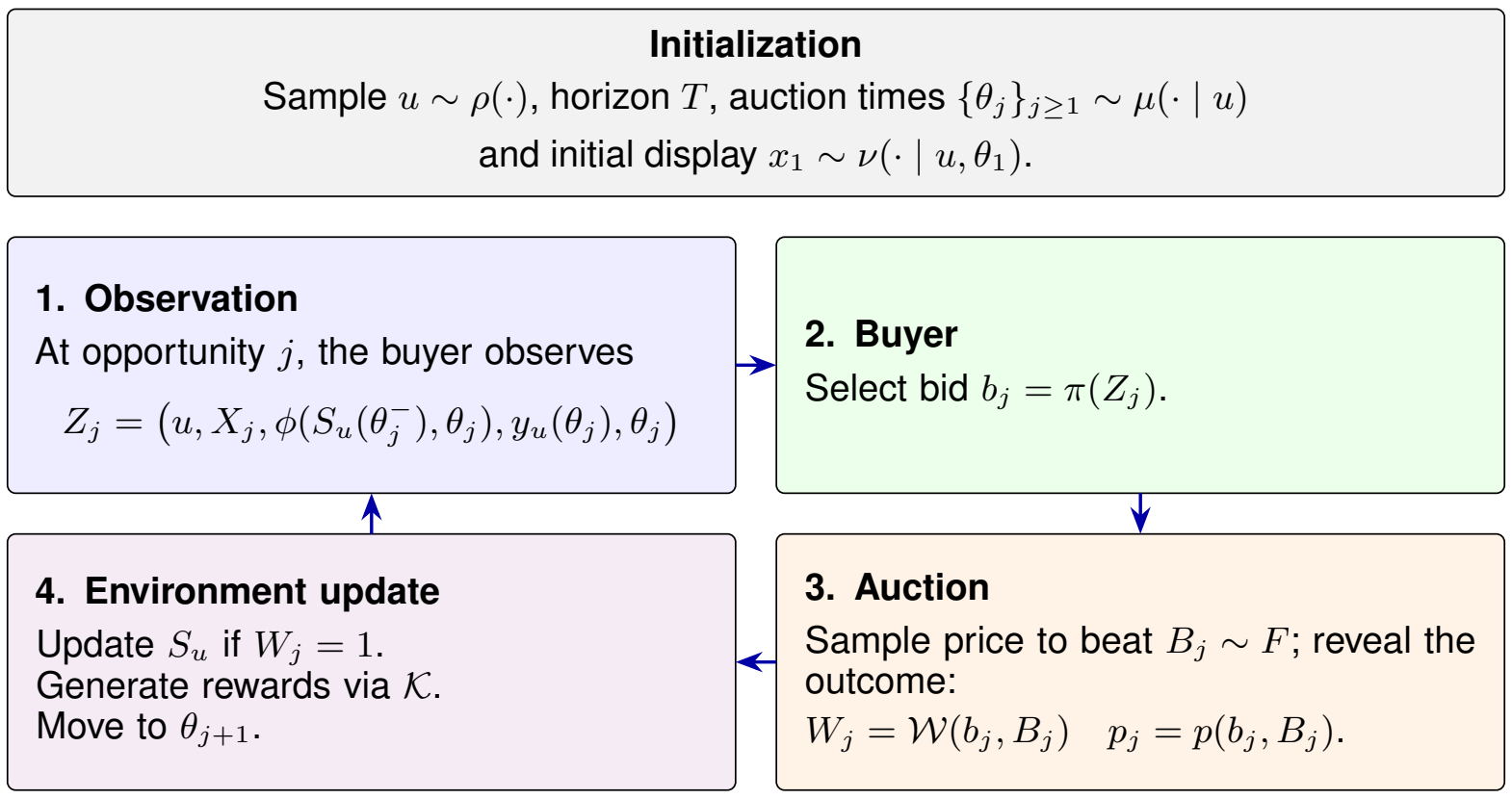

\paragraph{Information available to the buyer.}
Let
$\phi:\mathcal{S}\times[0,\infty)\to\mathcal{O}_S$ be a measurable observation map, where $\mathcal{O}_S$ is the space of reported carry-over summaries. Full observation corresponds to $\mathcal{O}_S=\mathcal{S}$ and $\phi(s,t)=s$. At an opportunity time $\theta_j$ on user $u$, we assume the buyer observes the candidate item feature $x_j$, the user identity $u$, the carry-over summary $\phi(S_u(\theta_j^-),\theta_j)$, the exogenous process value $y_u(\theta_j)$, and the time $\theta_j$ itself. We collect this information in the \textit{observable state}. 
\begin{equation*}
      Z_j =
  \bigl(
    u,X_j,\phi(S_u(\theta_j^-),\theta_j),
    y_u(\theta_j),\theta_j
  \bigr) \in
  \mathcal{U}\times\mathcal{X}\times\mathcal{O}_S
  \times\mathcal{Y}\times[0,\infty).
\end{equation*}
After each auction, the buyer observes the win indicator and payment. The scenario specifies whether the competing bid $B_j$ is also disclosed and how delayed rewards are reported. Policies may use current and past observations, past bids, feedback already received, and private randomization. The observation $Z_j$ need not be Markov. Writing $b_j=\pi(Z_j)$ restricts attention to policies using only the current observation. Figure \ref{fig:env-schema} summarizes the interaction process. 

\section{DynamAuction}
\label{sec:env}

\textsc{DynamAuction} implements the model introduced in the previous section as a Gymnasium environment. It allows researchers to evaluate bidding strategies under different auction dynamics through configurable parameters and provides a set of baseline policies for comparison.
Additional information is provided in Appendix~\ref{sec-appendix-implementation-details}.

\paragraph{Scenario configuration.}
In \textsc{DynamAuction}, a scenario defines the model parameters, including auction mechanism, reward kernel, carry-over context, competition distribution, campaign horizon, and arrival process. Scenarios can be saved and reused across experiments, making it easy to compare bidding strategies under the same conditions. Table~\ref{tab:presets} presents the predefined scenarios available in \textsc{DynamAuction}.

\paragraph{Gymnasium interface.}
\textsc{DynamAuction} follows the Gymnasium API~\citep{towers2024gymnasium} and can be used with RL libraries such as Stable-Baselines3~\citep{stable-baselines3}. We provide three wrappers to adapt the environment for RL training (Table~\ref{tab:wrappers}). The recommended order is: (i) \emph{oracle masking}, which removes expected-reward information that would not be available to a real bidder; (ii) \emph{observation flattening}, which converts the observations into a single vector for standard neural network policies; and (iii) \emph{bid clipping}, which limits bids to $[0,b_{\max}]$ during training for stable exploration.

\section{Model Predictive Control with Myopic Valuations}
\label{sec:mpc}

\subsection{Myopic-optimal valuation}

The planning problem introduced in Section~\ref{sec:model} can, in principle, be solved using dynamic programming. The optimal bid at time $t$ must account not only for the reward obtained by winning the current auction, but also for its impact on future rewards. As shown in \cite[Theorem 3.1]{10.1145/3447548.3467280}, this trade-off can be characterized using the Bellman value function $V^\star$, which represents the maximum expected cumulative reward achievable from a given state $z$. However, estimating $V^\star$ is challenging. The state space is high-dimensional, online exploration is costly, offline counterfactual estimation suffers from high variance, and the environment is non-stationary. Instead, we build on the notion of \textit{myopic-optimal valuation} introduced in \cite{bompaire2024fixed}. The myopic-optimal value $V_0$ is defined as the optimal bid for the current display under two assumptions: (i) the bidder cannot participate in any future auction on the same user, and (ii) the current auction is second-price. In our model, $V_0$ corresponds to the expected reward generated by the current display, obtained by integrating Equation~\eqref{eq:per-display-reward} over the full delay support. It can be estimated using a fixed continuation policy that places no further bids on the same user, thereby isolating the reward of the current display from future bidding decisions. Additional algorithmic details are provided in Appendix~\ref{app:algorithms}.

\subsection{Planning with Myopic-optimal Valuations}
\label{sec:mpc-on-top}
At each decision step, our Model Predictive Control (MPC) approach
compares candidate bids through Monte Carlo rollouts over a finite
planning horizon. It submits the bid with the highest estimated net
reward and replans at the next auction. An estimator $\widehat V_0$
of the myopic-optimal value provides the scale for both the candidate
bids and a fixed continuation policy. Rollout scores are computed
from simulated conversions and payments. Thus, the optimization
concerns a single multiplier for the current bid, while accounting
for its consequences for future opportunities.

\textit{Candidate bid.}
We express each candidate bid as a multiple of the estimated
myopic-optimal value:
\begin{equation}
  \label{eq:shading}
  b(\alpha;Z)=\alpha\widehat V_0(Z),
\end{equation}
where $\alpha\in\mathcal A_0\subset\mathbb R_{\geq0}$ is selected
from a finite set of candidate multipliers. In a second-price
auction, $\alpha=1$ bids the estimated myopic value, while
$\alpha<1$ allows the planner to reduce the bid when winning the
current auction has an adverse effect on future opportunities.

\textit{Simulating future auctions.}
At auction $i$, at time $\theta_i$, we generate $M$ rollouts from
copies of the current simulator state, retaining past wins and
pending conversions. The current observation $Z_i$ is held fixed,
while the remaining campaign duration and future events are
resampled according to the scenario model. Let $L_m$ denote the
last auction time in the complete sampled campaign, including the
current auction. The planning window contains the auction indices
$\mathcal N_H^{(m)}$ with
$\theta_i\leq\theta_k^{(m)}\leq\min(\theta_i+H,L_m)$,
where $H$ is measured in time units.
The current auction uses $b(\alpha;Z_i)$, and all subsequent
auctions in this window use $b(\alpha_{\mathrm{cont}};Z_k^{(m)})$,
with $\alpha_{\mathrm{cont}}$ fixed across candidates.
The continuation rule is fixed, but its bids depend on the
observations reached in each rollout. All candidates are evaluated
on the same sampled campaigns, using common random numbers for
competing bids, display features, and conversions, with draws
aligned by auction to reduce the variance of score differences.

\textit{Evaluating simulated trajectories.}
For each simulated win, the display's conversions are sampled
using its pre-win carry-over context. Let $C_{k,L_m}^{(m)}$ denote
the total value of these conversions occurring no later than
$L_m$. The contribution of auction $k$ to the rollout score is
\begin{equation}
  \label{eq:per-step-surrogate}
  g_k^{(m)}(\alpha)
  =\mathcal W\!\left(b_k^{(m)},B_k^{(m)}\right)
    C_{k,L_m}^{(m)}
   -p\!\left(b_k^{(m)},B_k^{(m)}\right),
\end{equation}
where $b_k^{(m)}=b(\alpha_k^{(m)};Z_k^{(m)})$,
$\alpha_i^{(m)}=\alpha$, and
$\alpha_k^{(m)}=\alpha_{\mathrm{cont}}$ for $k>i$.
The allocation and payment rules $\mathcal W$ and $p$ are those
defined in Section~\ref{sec:model}, with zero payment after a loss.
For a lost auction, we set $C_{k,L_m}^{(m)}=0$.

The planning horizon limits the auctions for which decisions are
simulated, not the delays of their conversions. A display won
within the window may therefore contribute conversions after
$\theta_i+H$, provided they occur before $L_m$, consistently with
the environment's reward reporting. Contributions from displays
won before auction $i$ are omitted from the score because they
do not depend on the candidate. The effects of simulated wins
on subsequent opportunities are captured through the evolution
of the carry-over context.

For each candidate, we average these returns over the $M$ rollouts:
\begin{equation}
  \label{eq:Q-hat}
  \widehat Q_i(\alpha)
  =\frac{1}{M}
   \sum_{m=1}^{M}
   \sum_{k\in\mathcal N_H^{(m)}}
   g_k^{(m)}(\alpha).
\end{equation}

\textit{Selecting the bid.}
We select the multiplier with the highest estimated net reward,
breaking ties in favor of the smallest multiplier:
\begin{equation}
  \label{eq:mpc-decision-main}
  \alpha_i^\star
  =\min\arg\max_{\alpha\in\mathcal A_0}\widehat Q_i(\alpha),
  \qquad
  b_i=b(\alpha_i^\star;Z_i).
\end{equation}
After submitting $b_i$, the bidder receives the auction feedback
and repeats the optimization at the next opportunity.
Pseudocode and implementation details are provided in
Appendix~\ref{app:algorithms}.

\section{Experiments}
\label{sec:experiments}
\paragraph{Algorithms.}
To illustrate the use of DynamAuction as a benchmark for bidding with dynamic valuations, we evaluate eight bidding policies, ranging from simple baselines to model-free reinforcement learning as well as our MPC approach. The policies are the following:

\begin{itemize}
    \item Constant: Bids a fixed amount $b^*$ selected by grid search.
    \item FiPla, introduced by~\cite{bompaire2024fixed} that bids the true myopic-optimal value $V_0(Z)$.
    \item Greedy Myopic, that estimates the marginal return of winning the current auction by comparing Monte Carlo rollouts that force a win or a loss, followed by a zero-bid continuation policy. The estimated difference is then used to determine the bid.
    \item Shading uses the same estimator as Greedy Myopic but scales the resulting bid by a constant multiplier $\alpha^*$ optimized through grid search.
    \item MPC, introduced in Section~\ref{sec:mpc}.
    \item Deep RL: We consider three model-free algorithms implemented with Stable-Baselines3~\cite{stable-baselines3}: PPO~\cite{schulman2017proximalpolicyoptimizationalgorithms}, SAC~\cite{haarnoja2018softactorcriticoffpolicymaximum}, and TD3~\cite{fujimoto2018addressingfunctionapproximationerror}.
\end{itemize}

All policies are evaluated on the same 1,000 test campaigns to ensure consistent comparisons. For each RL algorithm, we select hyperparameters using 200 validation campaigns and train three models with different random seeds. We report the average performance across these three runs. Further experimental details are provided in Appendix~\ref{app:experimental-details}.

\paragraph{Scenario.} We choose a scenario in which current bidding decisions have lasting effects on future rewards, making planning a central aspect of the task. This can be modelled by advertising fatigue: a user repeatedly exposed to the same advertisement over a short period may become less receptive to subsequent impressions, with an effect that persists throughout the campaign. The bidder must therefore balance the expected reward of winning the current auction against its effect on the value of subsequent opportunities, making the timing and spacing of wins relevant to campaign performance. This is a deliberate modeling choice for our benchmark. Users of DynamAuction are free to specify alternative carry-over dynamics, reward kernels, and scenario parameters to study different mechanisms or applications.

Formally, all main experiments use the
\texttt{survival-second-price-exponential} scenario, a second-price
(Vickrey) auction environment with a random campaign duration
$T\sim\mathrm{Exp}(1/\eta)$, where $\eta$ is the mean duration.
We consider $\eta=50$, the corresponding test campaigns contain
47.7 auctions on average. Auction opportunities arrive according
to a Poisson process of rate~1, and the competing bid is
$B\sim\mathrm{LogNormal}(-1,1)$. A user feature vector
$u\sim\mathcal{N}(0,I_3)$ is drawn once per campaign, and a display feature
vector $x\sim\mathcal{N}(0,I_3)$ is drawn at each auction. Performance is
reported as the mean episodic return over 1\,000 test campaigns.
The remaining parameters are listed in
Tables~\ref{tab:scenario-params-ab} and~\ref{tab:scenario-params-c}.

\textit{Carry-over context and observations.} To instantiate the commutative-monoid framework introduced in Section \ref{subsec:carry-over}, we take the carry-over
context to be the list of past wins:

\begin{equation}
    S_u(t^-)=\{t_i<t\}.
\end{equation}

At time $t$, the buyer only receives a scalar fatigue factor $\phi(S_u(t^-),t)$ computed from this history. The buyer therefore does not have access to the full carry-over context, making the scenario partially observable.
For $N\geq1$ past wins at times $t_1<\cdots<t_N<t$, define

\begin{equation}
      \phi(S_u(t^-),t)
  =\left[\prod_{i=2}^{N}
    \left(1-ce^{-\beta(t_i-t_{i-1})}\right)\right]
    \left(1-ce^{-\beta(t-t_N)}\right),
  \qquad c=0.9,\quad\beta=1.
\end{equation}

We use the empty-product convention and set
$\phi(\varnothing,t)=1$ before the first win. Each completed interval between consecutive wins contributes a fixed multiplicative penalty. Closely spaced wins cause a persistent reduction in subsequent conversion propensity within the campaign. If the current auction at time $t$ is won, its timestamp is added to the history after evaluating the current display's reward kernel, so it affects later displays through their carry-over context. The observation fields are described in Table~\ref{tab:obs-space}.

\textit{Delayed reward.}
There is no exogenous signal ($y_u\equiv0$) and no instantaneous reward
component ($\kappa=0$). For delay $\tau>0$, the reward kernel is

$$
      \mathcal{K}(u,x,S,0,t)(\mathrm{d}\tau)
  = \operatorname{clip}\!\left(
    \phi(S,t)(\mathbf{1}^\top u)(\mathbf{1}^\top x),
    0,1\right)\, e^{-\tau}\,\mathrm{d}\tau,
$$
Conditional on the realized campaign end $T$, the expected conversion
reward attributable to a display won at $t_i<T$ is
$$
  V_i^{\mathrm{oracle}}
  = \operatorname{clip}\!\left(
    \phi(S,t)(\mathbf{1}^\top u)(\mathbf{1}^\top x),
    0,1\right)
    \left(1-e^{-(T-t_i)}\right).
$$
We call this quantity the oracle value because it depends on the
realized end time $T$, which is unknown to the bidder. The oracle value is available to FiPla but hidden from the other agents (Table~\ref{tab:wrappers}).

\paragraph{Results.}

\begin{table}[t]
    \centering
    \caption{Mean episode reward ($\pm$ 95\% CI) over the 1\,000 test campaigns, and paired difference with Shading($\alpha = 0.4$, $M = 20$): its CI is computed from the per-campaign differences, which removes the variability between campaigns. RL agents: mean of their three trainings per campaign. Settings in Appendix~\ref{app:experimental-details}.}
    \label{tab:main-results}

    \small
    \setlength{\tabcolsep}{3.5pt}
    \begin{tabular}{lcc}
        \toprule
        Policy & Reward & $\Delta$ vs Shading \\
        \midrule
        MPC($H = 35$, $M = 60$, $\alpha_{\rm cont} = 0.4$) & $4.328 \pm 0.275$ & $+0.65 \pm 0.13$ \\
        MPC($H = 25$, $M = 60$, $\alpha_{\rm cont} = 0.4$) & $4.285 \pm 0.268$ & $+0.61 \pm 0.12$ \\
        MPC($H = 25$, $M = 80$) & $4.282 \pm 0.268$ & $+0.60 \pm 0.13$ \\
        MPC($H = 35$, $M = 60$) & $4.206 \pm 0.269$ & $+0.53 \pm 0.13$ \\
        MPC($H = 25$, $M = 60$) & $4.112 \pm 0.257$ & $+0.43 \pm 0.11$ \\
        MPC($H = 10$, $M = 60$) & $3.938 \pm 0.238$ & $+0.26 \pm 0.11$ \\
        Shading($\alpha = 0.4$, $M = 20$) & $3.680 \pm 0.234$ & --- \\
        PPO (mean of 3 trainings) & $2.965 \pm 0.182$ & $-0.71 \pm 0.12$ \\
        GreedyMyopic & $2.826 \pm 0.167$ & $-0.85 \pm 0.16$ \\
        FiPla & $2.821 \pm 0.160$ & $-0.86 \pm 0.16$ \\
        TD3 (mean of 3 trainings) & $2.006 \pm 0.121$ & $-1.67 \pm 0.15$ \\
        Constant($b = 0.11$) & $1.026 \pm 0.099$ & $-2.65 \pm 0.19$ \\
        SAC (mean of 3 trainings) & $0.707 \pm 0.119$ & $-2.97 \pm 0.24$ \\
        \bottomrule
    \end{tabular}
\end{table}

\begin{figure}[H]
  \centering
    \includegraphics[width=0.75\linewidth]{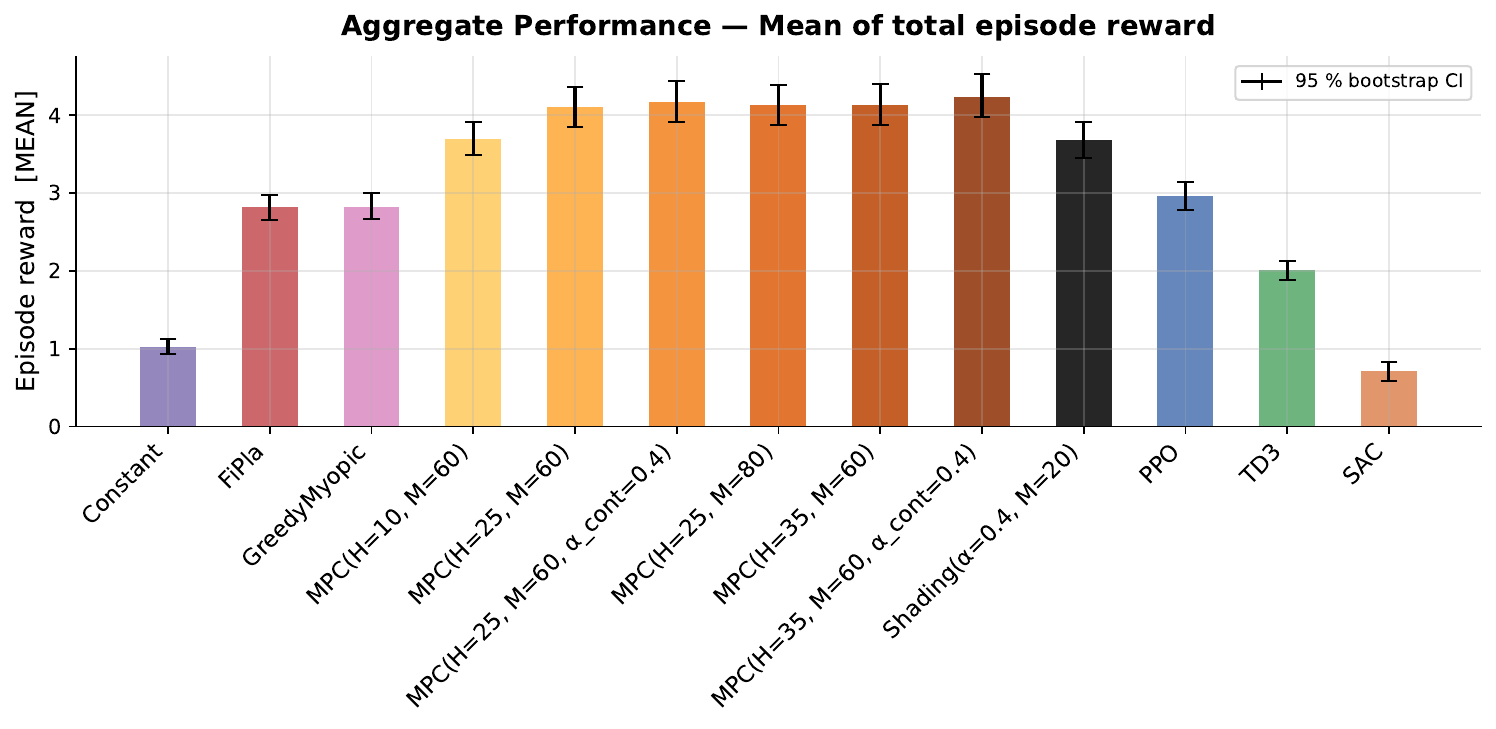}
  \caption{Mean episode reward per policy at $\eta = 50$ over the 1\,000 test campaigns.
    Error bars are 95\% bootstrap confidence intervals. RL agents: mean of their
    three trainings per campaign.}
  \label{fig:aggregate}
\end{figure}

Table~\ref{tab:main-results} and Figure~\ref{fig:aggregate} report the performance of the different policies. MPC outperforms the tuned Shading baseline when the planning horizon is sufficiently long ($H \geq 25$). The best configuration, MPC($H=35$, $M=60$, $\alpha_{\rm cont}=0.4$), achieves a mean episodic return of $4.33$, compared with $3.68$ for Shading and $2.82$ for FiPla. The gain over Shading is $+0.65 \pm 0.13$. With a shorter planning horizon ($H=10$), the improvement is smaller but remains positive ($+0.26 \pm 0.11$). Using a shaded continuation policy ($\alpha_{\rm cont}=0.4$) also improves MPC performance, although the gains are smaller. The deep RL algorithms perform worse than Shading on average. Their performance is also sensitive to training: PPO shows substantial variability across seeds, one of the three TD3 runs converges to a policy that never bids, and SAC wins considerably more auctions than FiPla but achieves a lower return. Additional experiments are reported in Appendix~\ref{app:additional-experiments}.

\section{Conclusion} 
\label{sec:conclusion}

We introduced a generic model for describing dynamic value in repeated auctions, along with DynamAuction, an RL environment for studying sequential bidding. The environment is challenging for deep reinforcement learning methods while accommodating settings that can be studied analytically, as in~\cite{heymann2024repeated}. On our benchmark, which illustrates the motivation for planning in this setting, simple shading and model-predictive control policies outperform the tested deep RL methods. 
Our design focuses on isolating the effects of dynamic value. Accordingly, we use synthetic data and leave budget and ROI constraints outside the scope of this work.

\begin{ack}
This work was supported in part by the French National Research Agency (ANR) through grants ANR-20-CE23-0007 and ANR-23-CE23-0002, and through the PEPR IA FOUNDRY project (ANR-23-PEIA-0003). It was also supported by Hi! PARIS and the ANR/France 2030 program (ANR-23-IACL-0005).
\end{ack}

\newpage

\bibliographystyle{abbrv}
\bibliography{ref.bib}

@inproceedings{weed2016online,
  title={Online learning in repeated auctions},
  author={Weed, Jonathan and Perchet, Vianney and Rigollet, Philippe},
  booktitle={Conference on Learning Theory},
  pages={1562--1583},
  year={2016},
  organization={PMLR}
}

@inproceedings{achddou2021efficient,
  title={Efficient algorithms for stochastic repeated second-price auctions},
  author={Achddou, Juliette and Capp{\'e}, Olivier and Garivier, Aur{\'e}lien},
  booktitle={Algorithmic Learning Theory},
  pages={99--150},
  year={2021},
  organization={PMLR}
}

@inproceedings{NIPS2017_0bed45bd,
 author = {Flajolet, Arthur and Jaillet, Patrick},
 booktitle = {Advances in Neural Information Processing Systems},
 editor = {I. Guyon and U. Von Luxburg and S. Bengio and H. Wallach and R. Fergus and S. Vishwanathan and R. Garnett},
 pages = {},
 publisher = {Curran Associates, Inc.},
 title = {Real-Time Bidding with Side Information},
 url = {https://proceedings.neurips.cc/paper_files/paper/2017/file/0bed45bd5774ffddc95ffe500024f628-Paper.pdf},
 volume = {30},
 year = {2017}
}

@inproceedings{10.5555/3618408.3619926,
author = {Wang, Qian and Yang, Zongjun and Deng, Xiaotie and Kong, Yuqing},
title = {Learning to bid in repeated first-price auctions with budgets},
year = {2023},
publisher = {JMLR.org},
booktitle = {Proceedings of the 40th International Conference on Machine Learning},
articleno = {1518},
numpages = {20},
location = {Honolulu, Hawaii, USA},
series = {ICML'23}
}

@inproceedings{10.1145/3033274.3084088,
author = {Balseiro, Santiago R. and Gur, Yonatan},
title = {Learning in Repeated Auctions with Budgets: Regret Minimization and Equilibrium},
year = {2017},
isbn = {9781450345279},
publisher = {Association for Computing Machinery},
address = {New York, NY, USA},
url = {https://doi.org/10.1145/3033274.3084088},
doi = {10.1145/3033274.3084088},
booktitle = {Proceedings of the 2017 ACM Conference on Economics and Computation},
pages = {609},
numpages = {1},
location = {Cambridge, Massachusetts, USA},
series = {EC '17}
}

@inproceedings{balseiro2020dual,
  title={Dual mirror descent for online allocation problems},
  author={Balseiro, Santiago and Lu, Haihao and Mirrokni, Vahab},
  booktitle={International Conference on Machine Learning},
  pages={613--628},
  year={2020},
  organization={PMLR}
}

@book{lattimore2020bandit,
  title={Bandit algorithms},
  author={Lattimore, Tor and Szepesv{\'a}ri, Csaba},
  year={2020},
  publisher={Cambridge University Press}
}

@article{balseiro2015repeated,
  title={Repeated auctions with budgets in ad exchanges: Approximations and design},
  author={Balseiro, Santiago R and Besbes, Omar and Weintraub, Gabriel Y},
  journal={Management Science},
  volume={61},
  number={4},
  pages={864--884},
  year={2015},
  publisher={INFORMS}
}

@inproceedings{10.1145/3038912.3052682,
author = {Balseiro, Santiago and Kim, Anthony and Mahdian, Mohammad and Mirrokni, Vahab},
title = {Budget Management Strategies in Repeated Auctions},
year = {2017},
isbn = {9781450349130},
publisher = {International World Wide Web Conferences Steering Committee},
address = {Republic and Canton of Geneva, CHE},
url = {https://doi.org/10.1145/3038912.3052682},
doi = {10.1145/3038912.3052682},
booktitle = {Proceedings of the 26th International Conference on World Wide Web},
pages = {15–23},
numpages = {9},
location = {Perth, Australia},
series = {WWW '17}
}

@article{castiglioni2022unifying,
  title={A unifying framework for online optimization with long-term constraints},
  author={Castiglioni, Matteo and Celli, Andrea and Marchesi, Alberto and Romano, Giulia and Gatti, Nicola},
  journal={Advances in Neural Information Processing Systems},
  volume={35},
  pages={33589--33602},
  year={2022}
}

@inproceedings{10.5555/3692070.3692294,
author = {Castiglioni, Matteo and Celli, Andrea and Kroer, Christian},
title = {Online learning under budget and ROI constraints via weak adaptivity},
year = {2024},
publisher = {JMLR.org},
booktitle = {Proceedings of the 41st International Conference on Machine Learning},
articleno = {224},
numpages = {25},
location = {Vienna, Austria},
series = {ICML'24}
}

@article{conitzer2022multiplicative,
  title={Multiplicative pacing equilibria in auction markets},
  author={Conitzer, Vincent and Kroer, Christian and Sodomka, Eric and Stier-Moses, Nicolas E},
  journal={Operations Research},
  volume={70},
  number={2},
  pages={963--989},
  year={2022},
  publisher={INFORMS}
}

@article{10.1145/3699824.3699838,
author = {Aggarwal, Gagan and Badanidiyuru, Ashwinkumar and Balseiro, Santiago R. and Bhawalkar, Kshipra and Deng, Yuan and Feng, Zhe and Goel, Gagan and Liaw, Christopher and Lu, Haihao and Mahdian, Mohammad and Mao, Jieming and Mehta, Aranyak and Mirrokni, Vahab and Leme, Renato Paes and Perlroth, Andres and Piliouras, Georgios and Schneider, Jon and Schvartzman, Ariel and Sivan, Balasubramanian and Spendlove, Kelly and Teng, Yifeng and Wang, Di and Zhang, Hanrui and Zhao, Mingfei and Zhu, Wennan and Zuo, Song},
title = {Auto-Bidding and Auctions in Online Advertising: A Survey},
year = {2024},
issue_date = {June 2024},
publisher = {Association for Computing Machinery},
address = {New York, NY, USA},
volume = {22},
number = {1},
url = {https://doi.org/10.1145/3699824.3699838},
doi = {10.1145/3699824.3699838},
journal = {SIGecom Exch.},
month = oct,
pages = {159–183},
numpages = {25}
}

@inproceedings{10.1145/3018661.3018702,
author = {Cai, Han and Ren, Kan and Zhang, Weinan and Malialis, Kleanthis and Wang, Jun and Yu, Yong and Guo, Defeng},
title = {Real-Time Bidding by Reinforcement Learning in Display Advertising},
year = {2017},
isbn = {9781450346757},
publisher = {Association for Computing Machinery},
address = {New York, NY, USA},
url = {https://doi.org/10.1145/3018661.3018702},
doi = {10.1145/3018661.3018702},
booktitle = {Proceedings of the Tenth ACM International Conference on Web Search and Data Mining},
pages = {661–670},
numpages = {10},
location = {Cambridge, United Kingdom},
series = {WSDM '17}
}

@inproceedings{10.1145/3269206.3271748,
author = {Wu, Di and Chen, Xiujun and Yang, Xun and Wang, Hao and Tan, Qing and Zhang, Xiaoxun and Xu, Jian and Gai, Kun},
title = {Budget Constrained Bidding by Model-free Reinforcement Learning in Display Advertising},
year = {2018},
isbn = {9781450360142},
publisher = {Association for Computing Machinery},
address = {New York, NY, USA},
url = {https://doi.org/10.1145/3269206.3271748},
doi = {10.1145/3269206.3271748},
booktitle = {Proceedings of the 27th ACM International Conference on Information and Knowledge Management},
pages = {1443–1451},
numpages = {9},
location = {Torino, Italy},
series = {CIKM '18}
}

@inproceedings{10.1145/3447548.3467199,
author = {He, Yue and Chen, Xiujun and Wu, Di and Pan, Junwei and Tan, Qing and Yu, Chuan and Xu, Jian and Zhu, Xiaoqiang},
title = {A Unified Solution to Constrained Bidding in Online Display Advertising},
year = {2021},
isbn = {9781450383325},
publisher = {Association for Computing Machinery},
address = {New York, NY, USA},
url = {https://doi.org/10.1145/3447548.3467199},
doi = {10.1145/3447548.3467199},
booktitle = {Proceedings of the 27th ACM SIGKDD Conference on Knowledge Discovery and Data Mining},
pages = {2993–3001},
numpages = {9},
location = {Virtual Event, Singapore},
series = {KDD '21}
}

@inproceedings{guo2024generative,
  title={Generative auto-bidding via conditional diffusion modeling},
  author={Guo, Jiayan and Huo, Yusen and Zhang, Zhilin and Wang, Tianyu and Yu, Chuan and Xu, Jian and Zheng, Bo and Zhang, Yan},
  booktitle={Proceedings of the 30th ACM SIGKDD Conference on Knowledge Discovery and Data Mining},
  pages={5038--5049},
  year={2024}
}

@inproceedings{10.1145/3746252.3761497,
author = {Peng, Yunshan and Shu, Wenzheng and Sun, Jiahao and Zeng, Yanxiang and Pang, Jinan and Bai, Wentao and Bai, Yunke and Liu, Xialong and Jiang, Peng},
title = {Expert-Guided Diffusion Planner for Auto-Bidding},
year = {2025},
isbn = {9798400720406},
publisher = {Association for Computing Machinery},
address = {New York, NY, USA},
url = {https://doi.org/10.1145/3746252.3761497},
doi = {10.1145/3746252.3761497},
booktitle = {Proceedings of the 34th ACM International Conference on Information and Knowledge Management},
pages = {5963–5970},
numpages = {8},
location = {Seoul, Republic of Korea},
series = {CIKM '25}
}

@article{bottou2013counterfactual,
  title={Counterfactual reasoning and learning systems: The example of computational advertising.},
  author={Bottou, L{\'e}on and Peters, Jonas and Qui{\~n}onero-Candela, Joaquin and Charles, Denis X and Chickering, D Max and Portugaly, Elon and Ray, Dipankar and Simard, Patrice and Snelson, Ed},
  journal={Journal of Machine Learning Research},
  volume={14},
  number={11},
  year={2013}
}

@inproceedings{10.1145/3447548.3467280,
author = {Bompaire, Martin and Gilotte, Alexandre and Heymann, Benjamin},
title = {Causal Models for Real Time Bidding with Repeated User Interactions},
year = {2021},
isbn = {9781450383325},
publisher = {Association for Computing Machinery},
address = {New York, NY, USA},
url = {https://doi.org/10.1145/3447548.3467280},
doi = {10.1145/3447548.3467280},
booktitle = {Proceedings of the 27th ACM SIGKDD Conference on Knowledge Discovery \& Data Mining},
pages = {75–85},
numpages = {11},
location = {Virtual Event, Singapore},
series = {KDD '21}
}

@article{bompaire2024fixed,
  title={Fixed point label attribution for real-time bidding},
  author={Bompaire, Martin and D{\'e}sir, Antoine and Heymann, Benjamin},
  journal={Manufacturing \& Service Operations Management},
  volume={26},
  number={3},
  pages={1043--1061},
  year={2024},
  publisher={INFORMS}
}

@article{lewis2022incrementality,
  title={Incrementality bidding and attribution},
  author={Lewis, Randall and Wong, Jeffrey},
  journal={arXiv preprint arXiv:2208.12809},
  year={2022}
}

@article{badanidiyuru2022incrementality,
  title={Incrementality bidding via reinforcement learning under mixed and delayed rewards},
  author={Badanidiyuru Varadaraja, Ashwinkumar and Feng, Zhe and Li, Tianxi and Xu, Haifeng},
  journal={Advances in Neural Information Processing Systems},
  volume={35},
  pages={2142--2153},
  year={2022}
}

@inproceedings{jeunen2023off,
  title={Off-policy learning-to-bid with auctiongym},
  author={Jeunen, Olivier and Murphy, Sean and Allison, Ben},
  booktitle={Proceedings of the 29th ACM SIGKDD Conference on Knowledge Discovery and Data Mining},
  pages={4219--4228},
  year={2023}
}

@inproceedings{10.1145/3696410.3714657,
author = {Khirianova, Alexandra and Solodneva, Ekaterina and Pudovikov, Andrey and Osokin, Sergey and Samosvat, Egor and Dorn, Yuriy and Ledovsky, Alexander and Zenkova, Yana},
title = {BAT: Benchmark for Auto-bidding Task},
year = {2025},
isbn = {9798400712746},
publisher = {Association for Computing Machinery},
address = {New York, NY, USA},
url = {https://doi.org/10.1145/3696410.3714657},
doi = {10.1145/3696410.3714657},
booktitle = {Proceedings of the ACM on Web Conference 2025},
pages = {2657–2667},
numpages = {11},
location = {Sydney NSW, Australia},
series = {WWW '25}
}

@article{su2024auctionnet,
  title={Auctionnet: A novel benchmark for decision-making in large-scale games},
  author={Su, Kefan and Huo, Yusen and Zhang, Zhilin and Dou, Shuai and Yu, Chuan and Xu, Jian and Lu, Zongqing and Zheng, Bo},
  journal={Advances in Neural Information Processing Systems},
  volume={37},
  pages={94428--94452},
  year={2024}
}

@inproceedings{10.1145/2648584.2648590,
author = {Liao, Hairen and Peng, Lingxiao and Liu, Zhenchuan and Shen, Xuehua},
title = {iPinYou Global RTB Bidding Algorithm Competition Dataset},
year = {2014},
isbn = {9781450329996},
publisher = {Association for Computing Machinery},
address = {New York, NY, USA},
url = {https://doi.org/10.1145/2648584.2648590},
doi = {10.1145/2648584.2648590},
booktitle = {Proceedings of the Eighth International Workshop on Data Mining for Online Advertising},
pages = {1–6},
numpages = {6},
location = {New York, NY, USA},
series = {ADKDD'14}
}

@article{zhang2014real,
  title={Real-time bidding benchmarking with ipinyou dataset},
  author={Zhang, Weinan and Yuan, Shuai and Wang, Jun and Shen, Xuehua},
  journal={arXiv preprint arXiv:1407.7073},
  year={2014}
}

@inproceedings{DiemertMeynet2017,
author = {{Diemert Eustache, Meynet Julien} and Galland, Pierre and Lefortier, Damien},
title={Attribution Modeling Increases Efficiency of Bidding in Display Advertising},
publisher = {ACM},
pages={To appear},
booktitle = {Proceedings of the AdKDD and TargetAd Workshop, KDD, Halifax, NS, Canada, August, 14, 2017},
year = {2017}
}

@book{CamachoBordons2013,
  title        = {Model Predictive Control},
  author       = {Camacho, Eduardo F. and Bordons, Carlos},
  edition      = {2},
  series       = {Advanced Textbooks in Control and Signal Processing},
  publisher    = {Springer London},
  year         = {2013},
  doi          = {10.1007/978-0-85729-398-5},
  isbn         = {978-0-85729-398-5},
}

@inproceedings{chua2018deepreinforcementlearninghandful,
  title     = {Deep Reinforcement Learning in a Handful of Trials using Probabilistic Dynamics Models},
  author    = {Chua, Kurtland and Calandra, Roberto and McAllister, Rowan and Levine, Sergey},
  booktitle = {Advances in Neural Information Processing Systems},
  year      = {2018},
  volume    = {31},
  note      = {},
}

@inproceedings{janner2021trustmodelmodelbasedpolicy,
  title     = {When to Trust Your Model: Model-Based Policy Optimization},
  author    = {Janner, Michael and Fu, Justin and Zhang, Marvin and Levine, Sergey},
  booktitle = {Advances in Neural Information Processing Systems},
  year      = {2019},
  volume    = {32},
  note      = {},
}

@inproceedings{hafner2020dreamcontrollearningbehaviors,
  title     = {Dream to Control: Learning Behaviors by Latent Imagination},
  author    = {Danijar Hafner and Timothy Lillicrap and Jimmy Ba and Mohammad Norouzi},
  booktitle = {International Conference on Learning Representations},
  year      = {2020},
  note      = {},
}

@article{Schrittwieser_2020,
   title={Mastering Atari, Go, chess and shogi by planning with a learned model},
   volume={588},
   ISSN={1476-4687},
   url={http://dx.doi.org/10.1038/s41586-020-03051-4},
   DOI={10.1038/s41586-020-03051-4},
   number={7839},
   journal={Nature},
   publisher={Springer Science and Business Media LLC},
   author={Schrittwieser, Julian and Antonoglou, Ioannis and Hubert, Thomas and Simonyan, Karen and Sifre, Laurent and Schmitt, Simon and Guez, Arthur and Lockhart, Edward and Hassabis, Demis and Graepel, Thore and Lillicrap, Timothy and Silver, David},
   year={2020},
   month=Dec, pages={604–609} }

@inproceedings{10.5555/3545946.3598810,
author = {Chen, Shuang and Xu, Qisen and Zhang, Liang and Jin, Yongbo and Li, Wenhao and Mo, Linjian},
title = {Model-Based Reinforcement Learning for Auto-bidding in Display Advertising},
year = {2023},
isbn = {9781450394321},
publisher = {International Foundation for Autonomous Agents and Multiagent Systems},
address = {Richland, SC},
booktitle = {Proceedings of the 2023 International Conference on Autonomous Agents and Multiagent Systems},
pages = {1560–1568},
numpages = {9},
location = {London, United Kingdom},
series = {AAMAS '23}
}

@article{stable-baselines3,
  author  = {Antonin Raffin and Ashley Hill and Adam Gleave and Anssi Kanervisto and Maximilian Ernestus and Noah Dormann},
  title   = {Stable-Baselines3: Reliable Reinforcement Learning Implementations},
  journal = {Journal of Machine Learning Research},
  year    = {2021},
  volume  = {22},
  number  = {268},
  pages   = {1-8},
  url     = {http://jmlr.org/papers/v22/20-1364.html}
}

@misc{schulman2017proximalpolicyoptimizationalgorithms,
      title={Proximal Policy Optimization Algorithms}, 
      author={John Schulman and Filip Wolski and Prafulla Dhariwal and Alec Radford and Oleg Klimov},
      year={2017},
      eprint={1707.06347},
      archivePrefix={arXiv},
      primaryClass={cs.LG},
      url={https://arxiv.org/abs/1707.06347}, 
}

@misc{fujimoto2018addressingfunctionapproximationerror,
      title={Addressing Function Approximation Error in Actor-Critic Methods}, 
      author={Scott Fujimoto and Herke van Hoof and David Meger},
      year={2018},
      eprint={1802.09477},
      archivePrefix={arXiv},
      primaryClass={cs.AI},
      url={https://arxiv.org/abs/1802.09477}, 
}

@misc{haarnoja2018softactorcriticoffpolicymaximum,
      title={Soft Actor-Critic: Off-Policy Maximum Entropy Deep Reinforcement Learning with a Stochastic Actor}, 
      author={Tuomas Haarnoja and Aurick Zhou and Pieter Abbeel and Sergey Levine},
      year={2018},
      eprint={1801.01290},
      archivePrefix={arXiv},
      primaryClass={cs.LG},
      url={https://arxiv.org/abs/1801.01290}, 
}

@inproceedings{10.1145/3543507.3583491,
author = {Feng, Zhe and Padmanabhan, Swati and Wang, Di},
title = {Online Bidding Algorithms for Return-on-Spend Constrained Advertisers},
year = {2023},
isbn = {9781450394161},
publisher = {Association for Computing Machinery},
address = {New York, NY, USA},
url = {https://doi.org/10.1145/3543507.3583491},
doi = {10.1145/3543507.3583491},
booktitle = {Proceedings of the ACM Web Conference 2023},
pages = {3550–3560},
numpages = {11},
location = {Austin, TX, USA},
series = {WWW '23}
}

@inproceedings{heymann2024repeated,
  title={Repeated Bidding with Dynamic Value},
  author={Heymann, Benjamin and Gilotte, Alexandre and Chan-Renous, R{\'e}mi},
  booktitle={International Conference on Web and Internet Economics},
  pages={548--563},
  year={2024},
  organization={Springer}
}

@article{towers2024gymnasium,
  title={Gymnasium: A Standard Interface for Reinforcement Learning Environments},
  author={Towers, Mark and Kwiatkowski, Ariel and Terry, Jordan and Balis, John U and De Cola, Gianluca and Deleu, Tristan and Goul{\~a}o, Manuel and Kallinteris, Andreas and Krimmel, Markus and KG, Arjun and others},
  journal={arXiv preprint arXiv:2407.17032},
  year={2024}
}

@inproceedings{Yang_2019, series={KDD ’19},
   title={Bid Optimization by Multivariable Control in Display Advertising},
   url={http://dx.doi.org/10.1145/3292500.3330681},
   DOI={10.1145/3292500.3330681},
   booktitle={Proceedings of the 25th ACM SIGKDD International Conference on Knowledge Discovery \& Data Mining},
   publisher={ACM},
   author={Yang, Xun and Li, Yasong and Wang, Hao and Wu, Di and Tan, Qing and Xu, Jian and Gai, Kun},
   year={2019}, pages={1966–1974},
   collection={KDD ’19} }

@article{heymann2026learning,
  title={Learning to Bid in Repeated Second-Price Auctions with Dynamic Values and Aggregated Feedback},
  author={Heymann, Benjamin and Sakhi, Otmane},
  journal={to be published at Neurips 2026},
  year={2026}
}

@inproceedings{heymann2019cost,
  title={Cost per action constrained auctions},
  author={Heymann, Benjamin},
  booktitle={Proceedings of the 14th Workshop on the Economics of Networks, Systems and Computation},
  pages={1--8},
  year={2019}
}
\newpage 
\appendix

\setcounter{tocdepth}{2} 
\addtocontents{toc}{\protect\setcounter{tocdepth}{2}}

\section*{Appendix}

\renewcommand{\contentsname}{}
\tableofcontents
\newpage

\section{Environment implementation and configuration}
\label{sec-appendix-implementation-details}
DynamAuction provides seven predefined scenarios (Table~\ref{tab:presets}), each of which can be customized using the parameters listed in Tables~\ref{tab:scenario-params-ab} and~\ref{tab:scenario-params-c}. Our main experiments use the \texttt{survival-second-price-exponential} scenario. Additional experiments consider immediate conversions and the saturation and recency scenarios (Appendix~\ref{app:additional-experiments}).

\begin{table}[H]
\centering
\footnotesize
\resizebox{\linewidth}{!}{%
\begin{tabular}{llll}
\toprule
\textbf{Preset ID} & \textbf{Kernel} & \textbf{Carry-over context}  \\
\midrule
\texttt{survival-second-price-exponential} & Survival & multiplicative fatigue  \\
\texttt{survival-recovering-second-price-exponential-burst-arrivals} & Survival & recovering multiplicative fatigue  \\
\texttt{survival-second-price-exponential-immediate} & Immediate survival & multiplicative fatigue \\
\texttt{saturation-second-price-lognormal} & Saturation & number of wins  \\
\texttt{recency-second-price-lognormal} & Recency & time of the last win  \\
\texttt{effective-frequency-second-price-lognormal} & Effective frequency & number of wins \\
\texttt{reminder-second-price-lognormal} & Reminder & time of the last win \\
\bottomrule
\end{tabular}}
\vspace{0.5em}
\caption{Built-in scenarios available in \textsc{DynamAuction}, with their reward kernels and carry-over contexts. Unless otherwise specified, all scenarios use second-price auctions, an exponentially distributed campaign duration with mean $\eta = 50$, and competing bids drawn from $\mathrm{LogNormal}(-1,1)$. See the \href{https://github.com/eugeniepatard/dynamauction}{GitHub repository} for the full scenario configurations}.
\label{tab:presets}
\end{table}

\subsection{Gymnasium Interface and Wrappers}
\label{sec:gym-appendix}

\textsc{DynamAuction} follows the Gymnasium API and can be used with Stable-Baselines3. For RL training, we apply three wrappers in order: remove the oracle expected reward from the observation, flatten the remaining observation fields into a vector, and clip bids to $[0,b_{\max}]$ (Table~\ref{tab:wrappers}).

\begin{table}[H]
  \centering
  \small
  \begin{tabular}{llp{6.2cm}}
    \toprule
    \textbf{Class} & \textbf{Type} & \textbf{Effect} \\
    \midrule
    \texttt{MaskExpectedRewardWrapper}
      & Observation
      & Removes \texttt{expected\_reward} from the observation dict.
        Prevents RL agents from accessing the oracle conversion value
        that \textsc{FiPla} bids directly, ensuring a fair comparison. \\[4pt]
    \texttt{FlatAuctionObsWrapper}
      & Observation
      & Concatenates all remaining \texttt{Dict} observation fields into
        a single 1-D \texttt{float32} vector, compatible with standard
        MLP policies. \\[4pt]
    \texttt{BidScaleWrapper}
      & Action
      & Clips the action space from $[0, 10^{12}]$ to $[0, b_{\max}]$
        (default $b_{\max} = 5.0$), restricting exploration to a
        practically relevant bid range. \\
    \bottomrule
  \end{tabular}
  \vspace{0.5em}
    \caption{Observation and action wrappers provided by \textsc{DynamAuction}.
    The recommended stacking order for RL training is top to bottom.}
    \label{tab:wrappers}
\end{table}

\subsection{Observation and Action Spaces}

Table~\ref{tab:obs-space} describes the observation fields provided by the environment. The user features in $Z_i$ (Section~\ref{sec:exo}) consist of an observed user vector $u_{\mathrm{obs}} \in \mathbb{R}^{d_o}$, with $d_o \leq d$, while the reward kernel depends on the full vector $u \in \mathbb{R}^d$. In all our experiments, the user features are fully observed ($d_o = d$).

\begin{table}[H]
\centering
\caption{Observation variables of the \textsc{DynamAuction} environment. Fields marked with $\dagger$ describe the previous auction and are excluded from policy inputs. The oracle value $V_0(Z_i)$ is used by \textsc{FiPla} but masked from RL observations.}
\label{tab:obs-space}
\small
\begin{tabular}{llp{6.5cm}}
\toprule
\textbf{Quantity} & \textbf{Dim.} & \textbf{Description} \\
\midrule
$\theta_{i}$                    & $1$      & Current auction time \\
$\phi(S_u(\theta^{-}_i), \theta_i)$ & $1$ & Carry-over summary just before the auction (Section~\ref{sec:model}): the context $S_u(\theta_i^-)$ itself when it is fully observed, or a scalar computed from it. \\
$i$                          & $1$      & Auction index within episode \\
$u_{\mathrm{obs}}$  & $d_o$    & Observable user features \\
$x_i$               & $d$      & Display feature vector \\
$y_u(\theta_{i})$                 & $1$      & Exogenous process value \\
$\mathbb{E}[\text{reward}\mid\text{win}_i,x_i,u]$ & $1$ & Oracle expected reward \\
$\tau_1, \ldots, \tau_{K}$ & $K$ & Times of the delayed conversions matured since $\theta_{i-1}$: the $K$ most recent ones, zero-padded if fewer ($K$ = \texttt{max\_history}, 1 in all presets) \\
$R_{[i-1,i]}$ & $1$ & Cumulative reward generated by the delayed conversions \\
\midrule
$B_i^\dagger$              & $1$      & Highest competing bid \\
$\mathbf{1}[\text{win}_i]^\dagger$ & $1$ & Win indicator \\
$p_i^\dagger$              & $1$      & Payment \\
\bottomrule
\end{tabular}
\end{table}

\subsection{Episode initialization and reward reporting}

At the start of each episode, the environment samples the campaign horizon $T$ and the auction times over $[0,T]$. Each step corresponds to an auction and advances the simulation to the next auction time. When an auction is won, the conversion times of the corresponding display are sampled. Delayed conversions are reported at the next auction, without identifying the display that generated them. Let $C_i^{\mathrm{now}}$ denote the number of immediate conversions from winning auction $i$, and $C^{\mathrm{delayed}}_{[i-1,i]}$ the number of delayed conversions occurring in $(\theta_{i-1},\theta_i]$, with $\theta_0=0$. The reward reported at step $i$ is

\begin{equation}
  \tilde r_i
  = v W_i C_i^{\mathrm{now}}
    - p_i
    + v C^{\mathrm{delayed}}_{[i-1,i]},
\end{equation}
where $v$ is the value of a conversion, $W_i$ indicates whether auction $i$ is won, and $p_i=p(b_i,B_i)$ is the payment, which is zero if the auction is lost. Unlike the expected reward $r_j(h)$ associated with a single display in Equation~\eqref{eq:per-display-reward}, the step reward $\tilde r_i$ may include conversions from several earlier wins. The episode ends after the last auction before $T$, and conversions occurring after this auction are not credited, even if they occur before $T$.

\section{Bidding Algorithms}
\label{app:algorithms}
We provide pseudocode and implementation details for MPC, followed by the GreedyMyopic and Shading baselines.

\subsection{MPC}

At auction $i$, which takes place at time $\theta_i$, the bidder
receives the observation $Z_i$. The learned estimator
$\widehat V_0(Z_i)$ predicts the conversion value of the current
display, before payment. MPC uses this estimate to construct
candidate bids, then compares them by simulating their consequences.

\paragraph{Candidate bids and future decisions.}
Let $\mathcal{A}_0$ be a finite set of nonnegative multipliers.
For each $\alpha \in \mathcal{A}_0$, the candidate bid is
$\max(0,\alpha\widehat V_0(Z_i))$. We evaluate this bid with $M$
simulated trajectories, called rollouts. Each rollout processes
the current auction and the following auctions up to time
$\theta_i+H$, where the planning horizon $H$ is measured in time units.

After the current auction, bids follow a fixed continuation policy.
At a later simulated auction $k$, this policy bids
$\max(0,\alpha_{\rm cont}\widehat V_0(Z_k))$, where $Z_k$ is the
observation reached in that rollout and $\alpha_{\rm cont}$ is a
fixed nonnegative multiplier. Thus, only the first multiplier
$\alpha$ is optimized; subsequent bids depend on the simulated
observations through $\widehat V_0$.

\paragraph{Simulated campaigns and shared random draws.}
Each rollout starts from a copy of the current simulator state,
including past wins and pending conversions. The current observation
is kept fixed, while future events are sampled anew. For the
exponential campaign durations used here, rollout $m$ draws a remaining
duration $T'_m \sim \mathrm{Exp}(1/\eta)$, where $\eta$ is the mean
campaign length. It then samples the future auction times and
exogenous events up to $\theta_i+T'_m$. We write $\theta_k^{(m)}$
for the auction times in this sampled campaign, with
$\theta_i^{(m)}=\theta_i$.

The same sampled campaign is used for every candidate $\alpha$.
Competing bids and future display features are drawn from separate
random streams, and conversion draws are indexed by auction.
This keeps the random inputs at auction $k$ aligned across
candidates, even when earlier auction outcomes differ. The conversion
outcomes can still differ, because their probabilities depend on the
history of wins. These common random numbers are used to reduce the
variance of differences between candidate returns.

\paragraph{Scoring a rollout.}
For each candidate $\alpha$, rollout $m$ simulates the current auction and subsequent auctions up to the planning horizon $H$. Its return is the net reward of the displays won during this window:
\[
R^{(m)}(\alpha)
=
\sum_{k=i}^{k_{\mathrm{end}}^{(m)}}
\mathcal{W}\bigl(b_k^{(m)}, B_k^{(m)}\bigr)
\Bigl(C_k^{(m)} - p\bigl(b_k^{(m)}, B_k^{(m)}\bigr)\Bigr),
\qquad
b_k^{(m)} = \alpha_k^{(m)}\,\widehat V_0\bigl(Z_k^{(m)}\bigr),
\]
where $\alpha_i^{(m)} = \alpha$ and $\alpha_k^{(m)} = \alpha_{\rm cont}$ for $k > i$. Here, $k_{\mathrm{end}}^{(m)}$ is the last auction within the planning horizon, and $C_k^{(m)}$ is the value of the conversions generated by the display won at auction $k$ that occur before the last auction of the simulated campaign, at time $t^{(N_m)}$.

Importantly, the planning horizon limits which auctions are simulated, not which conversions are counted. A display won within the planning horizon may generate conversions after $H$, and these are included in its return as long as they occur before $t^{(N_m)}$, consistently with the environment's reward reporting. Conversions from displays won before auction $i$ are omitted because they do not depend on the candidate $\alpha$.

The estimated return of each candidate is then
\[
\widehat Q(\alpha) = \frac{1}{M}\sum_{m=1}^{M}R^{(m)}(\alpha).
\]
MPC selects the candidate with the largest mean score, breaking ties
in favor of the smallest multiplier. It submits this bid and repeats
the procedure at the next real auction.
Algorithm~\ref{alg:mpc} gives the steps.

\begin{algorithm}[H]
\caption{MPC: selecting the bid at auction $i$}
\label{alg:mpc}
\begin{algorithmic}[1]
\Require Current simulator state, observation $Z_i$, auction time $\theta_i$
\Require Estimator $\widehat V_0$, multipliers $\mathcal{A}_0$,
         continuation multiplier $\alpha_{\rm cont}$,
         horizon $H$, number of rollouts $M$

\State $v_i \leftarrow \widehat V_0(Z_i)$

\For{$m=1,\ldots,M$}
    \State Draw a remaining duration $T'_m \sim \mathrm{Exp}(1/\eta)$
    \State Sample future auctions and exogenous events up to $\theta_i+T'_m$
    \State $L_m \leftarrow$ last auction time in the sampled campaign,
           including $\theta_i$
    \State $K_m \leftarrow$ last auction index with
           $\theta_k^{(m)} \leq \theta_i+H$
    \State Store the shared random streams for competing bids,
           display features and conversions
\EndFor

\For{each $\alpha \in \mathcal{A}_0$}
    \State $\widehat Q(\alpha) \leftarrow 0$

    \For{$m=1,\ldots,M$}
        \State Copy the current simulator state,
               keeping past wins and pending conversions
        \State Load rollout $m$'s auction schedule and exogenous events;
               reset its random streams
        \State $R \leftarrow 0$

        \For{$k=i,\ldots,K_m$}
            \State Advance the copy to auction $k$
                   and read its observation $Z_k^{(m)}$

            \If{$k=i$}
                \State $b \leftarrow \max(0,\alpha v_i)$
            \Else
                \State $b \leftarrow
                       \max(0,\alpha_{\rm cont}
                       \widehat V_0(Z_k^{(m)}))$
            \EndIf

            \State Determine the outcome and payment $p$ using $b$
                   and the competing bid of rollout $m$ at auction $k$

            \If{the auction is won}
                \State Sample and schedule the display's conversions
                       using its pre-win context and the draws for $(m,k)$
                \State $C \leftarrow$ total value of these conversions
                       occurring no later than $L_m$
                \State $R \leftarrow R+C-p$
            \EndIf

            \State Update the simulator copy with the auction outcome
        \EndFor

        \State $\widehat Q(\alpha)
               \leftarrow \widehat Q(\alpha)+R/M$
    \EndFor
\EndFor

\State $\alpha^* \leftarrow
       \min\arg\max_{\alpha\in\mathcal{A}_0}\widehat Q(\alpha)$
\State \Return $\max(0,\alpha^*v_i)$
\end{algorithmic}
\end{algorithm}

\paragraph{Special cases.}
With $\mathcal{A}_0=\{1\}$, MPC returns the bid
$\max(0,\widehat V_0(Z_i))$; using the oracle $V_0$ gives the
\textsc{FiPla} bid. With $H=0$, each rollout evaluates only the
current auction, including its delayed conversions credited before
$L_m$. In a second-price auction, $\alpha=1$ is optimal in expectation
if the valuation is exact for these credited conversions and
$1\in\mathcal{A}_0$. A learned valuation and a finite number of rollouts
can lead to a different choice.

\paragraph{Computational cost.}
Each decision evaluates $|\mathcal{A}_0|M$ rollouts.
The number of auctions processed in each rollout depends on $H$,
the arrival process and the sampled campaign duration.
The rollouts use the simulator's dynamics directly;
$\widehat V_0$ is the learned component used to set bids.

\subsection{GreedyMyopic}

GreedyMyopic estimates the conversion value of winning the current
auction, assuming that no further auctions will be won. Let $\pi_0$
denote the continuation policy that bids zero at every subsequent
auction.

At auction $i$, we run $M$ pairs of simulations from the current
simulator state. In each pair, one simulation forces a win at
auction $i$ and the other forces a loss. Both then follow $\pi_0$
until the last auction of the simulated campaign. They share the
same remaining duration $T'_m \sim \mathrm{Exp}(1/\eta)$, future
auction times and exogenous events, and start with the same noise
seed $\omega_m$.

Let $R_{\mathrm{win}}^{(m)}$ and $R_{\mathrm{loss}}^{(m)}$ denote
the total conversion value credited in these two simulations.
Payments are excluded, since the quantity being estimated is the
display's value before payment. As in the environment, conversions
occurring after the last auction are not credited. The bid is
\[
b_i = \hat\Delta
= \max\!\left(
    0,\;
    \frac{1}{M}\sum_{m=1}^{M}
    \left(R_{\mathrm{win}}^{(m)}
          - R_{\mathrm{loss}}^{(m)}\right)
  \right).
\]
The paired simulations use common random numbers to cancel the
contributions of previously won displays, as detailed in
Remark~\ref{rem:crn}.

\begin{algorithm}[H]
\caption{GreedyMyopic: selecting the bid at auction $i$}
\label{alg:greedy}
\begin{algorithmic}[1]
\Require Current simulator state at auction $i$ (observation $Z_i$),
         zero-bid continuation policy $\pi_0$, number of pairs $M$
\Ensure Bid $b_i \geq 0$

\State $S_{\mathrm{win}} \leftarrow 0$;
       \quad $S_{\mathrm{loss}} \leftarrow 0$

\For{$m=1,\ldots,M$}
    \State Draw a scenario seed and a noise seed $\omega_m$
    \State Use the scenario seed to sample
           $T'_m \sim \mathrm{Exp}(1/\eta)$,
           future auction times and exogenous events

    \State \textbf{Forced win:}
    \State Copy the current simulator state,
           keeping past wins and pending conversions
    \State Load the sampled campaign and initialize
           the noise generator with $\omega_m$
    \State Force auction $i$ to be won
    \State Follow $\pi_0$ through the last auction
           of the sampled campaign
    \State $R_{\mathrm{win}}^{(m)} \leftarrow$
           total conversion value credited in this simulation

    \State \textbf{Forced loss:}
    \State Make a fresh copy of the same current simulator state
    \State Load the same sampled campaign and reset
           the noise generator to $\omega_m$
    \State Force auction $i$ to be lost
    \State Follow $\pi_0$ through the last auction
           of the sampled campaign
    \State $R_{\mathrm{loss}}^{(m)} \leftarrow$
           total conversion value credited in this simulation

    \State $S_{\mathrm{win}} \leftarrow
           S_{\mathrm{win}}+R_{\mathrm{win}}^{(m)}$
    \State $S_{\mathrm{loss}} \leftarrow
           S_{\mathrm{loss}}+R_{\mathrm{loss}}^{(m)}$
\EndFor

\State \Return
       $\max\!\left(0,\;
       \dfrac{S_{\mathrm{win}}-S_{\mathrm{loss}}}{M}\right)$
\end{algorithmic}
\end{algorithm}

\begin{remark}[Common Random Numbers]
\label{rem:crn}The scenario seed fixes the remaining campaign duration, auction times and exogenous events. The noise seed $\omega_m$ initializes the single generator used for competing bids, display features and conversions. The forced-win simulation draws conversions for the current display, which shifts the subsequent draws of this generator. These later draws do not affect conversion rewards because $\pi_0$ wins no further auctions. Conversions already pending at auction $i$ are copied unchanged into both simulations and are credited at the same auction times. Their contributions therefore cancel in $R_{\mathrm{win}}^{(m)}-R_{\mathrm{loss}}^{(m)}$, leaving the conversion value credited to the current display. This value remains random, and $\hat\Delta$ averages it over the $M$ pairs.
\end{remark}

\subsection{Shading}
Shading uses the same value estimate $\hat\Delta$ as GreedyMyopic
(Algorithm~\ref{alg:greedy}), but multiplies it by a fixed factor
$\alpha \in (0,1]$. This reduces the bid to account for the effect
of winning the current auction on the value of future opportunities.
When $\alpha = 1$, Shading is equivalent to GreedyMyopic.
\begin{algorithm}[H]
\caption{Shading($\alpha$) — bid at auction $i$}
\label{alg:shading}
\begin{algorithmic}[1]
\Require current state $Z_i$, shading factor $\alpha \in (0,1]$,
         GreedyMyopic estimator (Algorithm~\ref{alg:greedy})
\Ensure bid $b_i \geq 0$
\State $\hat\Delta \leftarrow \text{GreedyMyopic}(Z_i)$
\State \Return $\max(0,\; \alpha \cdot \hat\Delta)$
\end{algorithmic}
\end{algorithm}

\section{Experimental details}
\label{app:experimental-details}

\subsection{Scenario configuration}
Tables~\ref{tab:scenario-params-ab} and~\ref{tab:scenario-params-c}
list the configurable components of \textsc{DynamAuction} and their
preset values. The main experiments use this preset with
$\eta = 50$.

\begin{table}[H]
\centering
\caption{Configurable components of a \textsc{DynamAuction} scenario: auction, horizon, competition, arrival, user features, and display features. A dash (---) indicates inactive components.}
\label{tab:scenario-params-ab}
\small
\setlength{\tabcolsep}{4pt}

{\raggedright\textbf{(a) Auction, horizon, competition, arrival}\par}
\vspace{0.3em}
\begin{tabularx}{\textwidth}{l l X l}
\toprule
\textbf{Component} & \textbf{Parameter} & \textbf{Options / Type} & \textbf{Preset} \\
\midrule
Auction & mechanism & \texttt{second-price}, \texttt{first-price} & second-price \\
\midrule
Horizon & type & \texttt{exponential}, \texttt{fixed} & exponential \\
& $\eta$ (\texttt{exponential}) & $\mathbb{R}_{+}$, mean campaign length & 50 \\
& $T$ (\texttt{fixed}) & $\mathbb{R}_{+}$, campaign length & --- \\
\midrule
Competition & distribution & \texttt{lognormal}, \texttt{uniform}, \texttt{exponential}, \texttt{constant} & lognormal \\
& $\mu_B$, $\sigma_B$ (\texttt{lognormal}) & $\mathbb{R}$, $\mathbb{R}_{+}$ & $-1.0$, $1.0$ \\
& low, high (\texttt{uniform}) & $\mathbb{R}_{+}$ & --- \\
& scale (\texttt{exponential}) & $\mathbb{R}_{+}$ & --- \\
& value (\texttt{constant}) & $\mathbb{R}_{+}$, the same competing bid at every auction & --- \\
\midrule
Arrival & process & \texttt{cstpoisson}, \texttt{pwpoisson}, \texttt{decaypoisson} & \texttt{cstpoisson} \\
& rate $\mu$ (\texttt{cstpoisson}) & $\mathbb{R}_{+}$ & 1.0 \\
& breakpoints, intensities (\texttt{pwpoisson}) & piecewise-constant rate: lists in $\mathbb{R}_{+}$ & --- \\
& rate, period, factor (\texttt{decaypoisson}) & rate $\times$ factor$^{\lfloor t/\text{period} \rfloor}$, in $\mathbb{R}_{+}$ & --- \\
\bottomrule
\end{tabularx}

\vspace{1em}

{\raggedright\textbf{(b) User and display features}\par}
\vspace{0.3em}
\begin{tabularx}{\textwidth}{l l X l}
\toprule
\textbf{Component} & \textbf{Parameter} & \textbf{Options / Type} & \textbf{Preset} \\
\midrule
User features & sampler type & \texttt{uniform}, \texttt{normal}, \texttt{fixed} & normal \\
& full dim $d$ & $\mathbb{N}$ & 3 \\
& observable dim $d_o \leq d$ & $\mathbb{N}$ & 3 \\
& mean $\mu_u$ & $\mathbb{R}$ & 0.0 \\
& std $\sigma_u$ & $\mathbb{R}_{+}$ & 1.0 \\
& feature vector $u$ & $\mathbb{R}^d$ & --- \\
\midrule
Display features & sampler type & \texttt{iid-uniform}, \texttt{catalogue}, \texttt{fixed}, \texttt{normal} & normal \\
& IIDUniformSampler & $x_i \sim \mathcal{U}([0,1]^d)$ & --- \\
& CatalogueSampler & $x_i \sim \mathcal{C}$ & --- \\
& FixedSampler & $x_i = x_0$ & --- \\
& NormalSampler & $x_i \sim \mathcal{N}(\mu,\sigma^2 I)$ & $\mu = 0$, $\sigma = 1$ \\
\bottomrule
\end{tabularx}
\end{table}


\begin{table}[H]
\centering
\caption{Configurable components of a \textsc{DynamAuction} scenario: value model, context, exogenous process, and conversion generation.}
\label{tab:scenario-params-c}
\small
\setlength{\tabcolsep}{4pt}

{\raggedright\textbf{(c) Value model, context, and exogenous process}\par}
\vspace{0.3em}
\noindent
\begin{tabularx}{\textwidth}{@{} l >{\raggedright\arraybackslash}X >{\raggedright\arraybackslash}X l @{}}
\toprule
\textbf{Component} & \textbf{Parameter} & \textbf{Options / Type} & \textbf{Preset} \\
\midrule
Value kernel 
  & type 
  & \texttt{survival}, \texttt{immediate-survival}, \texttt{saturation}, \texttt{recency}, \texttt{competitor-extinction}, \texttt{effective-frequency}, \texttt{reminder}, \texttt{mixed} 
  & survival \\
  & $\mathbf{w}_u, \mathbf{w}_x$ (\texttt{survival}, \texttt{immediate-survival}; optional relevance for \texttt{effective-frequency}, \texttt{reminder}) 
  & $\mathbb{R}^d$ 
  & $\mathbf{1}, \mathbf{1}$ \\
  & delay rate $\delta$ (all but \texttt{mixed}) 
  & $\mathbb{R}_{+}$ 
  & 1.0 \\
  & base rate, saturation scale (\texttt{saturation}) 
  & $\mathbb{R}_{+}$ 
  & --- \\
  & recency rate (\texttt{recency}) 
  & $\mathbb{R}_{+}$ 
  & --- \\
  & base rate (\texttt{competitor-extinction}) 
  & $\mathbb{R}_{+}$ 
  & --- \\
  & base rate, wear-in midpoint and slope, saturation scale (\texttt{effective-frequency}) 
  & $\mathbb{R}_{+}$ 
  & --- \\
  & $p_{\mathrm{cold}}$, $p_{\mathrm{peak}}$, peak time (\texttt{reminder}) 
  & $[0, 1]$, $[0, 1]$, $\mathbb{R}_{+}$ 
  & --- \\
  & base kernel, $\kappa_{\mathrm{base}}$ (\texttt{mixed}) 
  & kernel type, $\mathbb{R}_{+}$ 
  & --- \\
\midrule
Carry-over $S_u$ and decay map $\gamma$ 
  & type 
  & \texttt{multiplicative}, \texttt{recovering-multiplicative}, \texttt{additive}, \texttt{last-won-time}, \texttt{time-since-last-win}, \texttt{constant} 
  & multiplicative \\
  & decay $c$ (\texttt{multiplicative}, \texttt{recovering-multiplicative}) 
  & $(0,1)$ 
  & 0.9 \\
  & recovery rate $\beta$ (\texttt{multiplicative}, \texttt{recovering-multiplicative}) 
  & $\mathbb{R}_{+}$ 
  & 1.0 \\
  & initial value (\texttt{multiplicative}, \texttt{additive}, \texttt{last-won-time}, \texttt{time-since-last-win}) 
  & $\mathbb{R}$, context before the first win 
  & 1.0 \\
  & value (\texttt{constant}) 
  & $\mathbb{R}$ 
  & --- \\
\midrule
Exogenous process 
  & type 
  & \texttt{trivial} ($y(t) = 0$), \texttt{competitor-click} 
  & trivial \\
  & click rate (\texttt{competitor-click}) 
  & $\mathbb{R}_{+}$, Poisson rate of competitor clicks; $y(t)$ is the time of the last click 
  & --- \\
\bottomrule
\end{tabularx}

\vspace{0.8em}

{\raggedright\textbf{(d) Conversion generation and delay modeling}\par}
\vspace{0.3em}
\begin{tabularx}{\textwidth}{l l X l}
\toprule
\textbf{Component} & \textbf{Parameter} & \textbf{Options / Type} & \textbf{Preset} \\
\midrule
\textbf{Conversion probe} & type
& \texttt{poisson}, \texttt{bernoulli}, \texttt{null}
& bernoulli \\
\cmidrule(l){2-4}
\quad Poisson probe & sampling method
& Ogata thinning process
& --- \\
& event generation
& point process over horizon
& --- \\
\cmidrule(l){2-4}
\quad Bernoulli probe & conversion gate
& Bernoulli + exponential delay
& \checkmark \\
& max conversions
& 0 or 1 per display
& --- \\
\quad Null probe & rule
& no delayed conversion (immediate kernels)
& --- \\
\midrule
\textbf{Delay sampler} & type
& \texttt{single}, \texttt{multi}
& single \\
\cmidrule(l){2-4}
\quad Single sampler & rule
& at most one conversion possible per display
& \checkmark \\
\quad Multi sampler & rule
& multiple conversions allowed
& --- \\
\midrule
Display history & \texttt{max\_history} & $\mathbb{N}$ & 1 \\
\bottomrule
\end{tabularx}

\end{table}

\subsection{Evaluation protocol}

All policies are evaluated on the same 1\,000 test campaigns, generated with environment seeds 0--999. This allows us to compare policies on identical campaigns, rather than on independently sampled ones. For stochastic policies, campaign $s$ uses the policy seed $s+1000$.

Training and model selection use different campaigns. The MPC valuation head is trained on 1\,000 \textsc{FiPla} campaigns (seeds 10\,000--10\,999), while the RL agents generate their training campaigns from their own random streams. Intermediate evaluations during RL training use seeds $\geq 10\,000$. RL hyperparameters and checkpoints are selected using 200 validation campaigns (seeds 20\,000--20\,199). Each selected RL configuration is then trained with three different random seeds.

We report mean rewards with 95\% confidence intervals, written as the mean $\pm 1.96\,\hat\sigma/\sqrt{n}$, where $\hat\sigma$ is the sample standard deviation over the $n$ test campaigns. For comparisons between policies $A$ and $B$, we compute the reward difference $R_A(s)-R_B(s)$ on each campaign $s$ and use the standard deviation of these paired differences to construct the confidence interval. For RL policies, $R(s)$ is the average reward of the three trained models on campaign $s$. Since both policies face the same campaigns, pairing reduces the variability due to differences in campaign length and user characteristics, yielding narrower confidence intervals for their performance differences.

\subsection{Hyperparameters and training}

We report the configurations used in the experiments and how their parameters were selected.

\paragraph{MPC.}
MPC evaluates candidate bids using the multiplier grid
$\mathcal{A}_0 = \{0.1, 0.2, \dots, 1.0\}$, with ties resolved in favor
of the smallest multiplier. We consider planning horizons
$H \in \{10, 25, 35\}$, $M \in \{60, 80\}$ rollouts per candidate,
and continuation multipliers $\alpha_{\rm cont} \in \{1, 0.4\}$,
where $0.4$ is the multiplier selected for Shading. Unless otherwise
specified, $\alpha_{\rm cont} = 1$ in all tables and figures.

As described in Algorithm~\ref{alg:mpc}, candidate bids are compared
using common random numbers for competing bids, display features,
and conversions. Each rollout is scored by the net reward of the
displays won within the planning horizon, including their conversions
occurring before the last auction of the simulated campaign.

MPC uses a learned valuation head $\widehat V_0$ to estimate the
oracle value $V_0$ from observations that exclude $V_0$ itself.
We train this estimator on 1\,000 \textsc{FiPla} campaigns,
corresponding to 49\,358 auctions at $\eta = 50$, with 15\% of the
data held out for validation. Training uses Adam with a learning
rate of $10^{-3}$, a batch size of 256, and 500 epochs.

\begin{figure}[H]
\centering
\includegraphics[width=\linewidth]{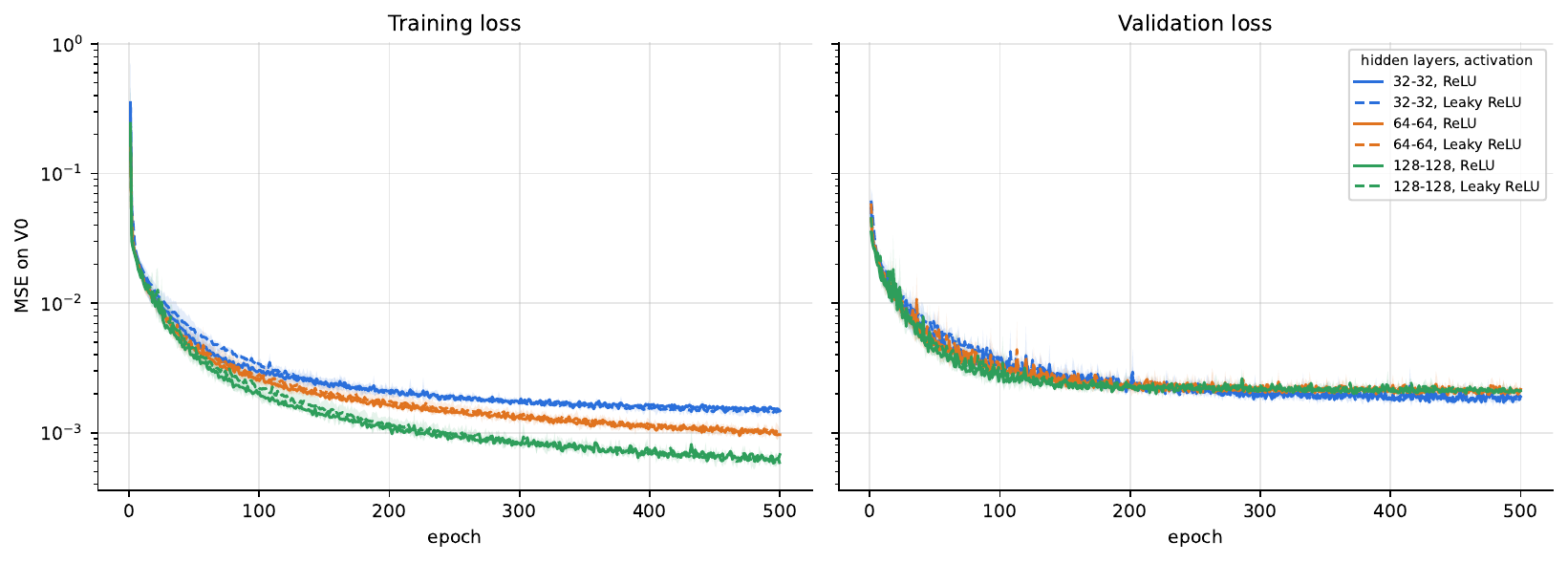}

\caption{Training and validation mean squared error (MSE) of MLPs used to estimate the oracle value $V_0$. We compare three network sizes (32-32, 64-64, and 128-128 hidden units) with ReLU and Leaky ReLU activations. All models are trained and evaluated on the same data split from the main scenario. Curves show the mean over three random initializations, with shaded areas indicating the minimum and maximum values. MPC uses the 32-32 ReLU network.}
\label{fig:mlp-curves}
\end{figure}

We compare MLPs with two hidden layers of 32, 64, or 128 units,
using ReLU or Leaky ReLU activations (Figure~\ref{fig:mlp-curves}
and Table~\ref{tab:mlp-arch}). Although the larger networks achieve
lower training MSE (0.00097 and 0.00068, compared with 0.00147
for 32-32 ReLU), their test errors are higher (0.00208 and
0.00209, compared with 0.00184). Leaky ReLU performs similarly
to ReLU (test MSE 0.00181 for 32-32 Leaky ReLU). We therefore
use the 32-32 ReLU network, which has the lowest test error
among the ReLU architectures and requires only 1\,505 parameters,
compared with 5\,057 and 18\,305 for the larger networks.
Its smaller size also reduces the cost of the repeated predictions
made during MPC rollouts.

\begin{table}[H]
\centering
\caption{MLP valuation heads of different sizes and activations (mean over 3 initialization seeds). Test: 200 new \textsc{FiPla} campaigns.}
\label{tab:mlp-arch}
\footnotesize
\begin{tabular}{llcccccc}
\toprule
hidden layers & activation & parameters & train MSE & validation MSE & test MSE & test corr. & minutes \\
\midrule
32-32 & relu & 1505 & 0.00147 & 0.00190 & 0.00184 & 0.988 & 5.7 \\
32-32 & leaky relu & 1505 & 0.00145 & 0.00186 & 0.00181 & 0.988 & 5.7 \\
64-64 & relu & 5057 & 0.00097 & 0.00216 & 0.00208 & 0.986 & 6.0 \\
64-64 & leaky relu & 5057 & 0.00101 & 0.00216 & 0.00227 & 0.986 & 6.0 \\
128-128 & relu & 18305 & 0.00068 & 0.00211 & 0.00209 & 0.987 & 6.7 \\
128-128 & leaky relu & 18305 & 0.00057 & 0.00207 & 0.00208 & 0.987 & 6.5 \\
\bottomrule
\end{tabular}
\end{table}

\begin{figure}[H]
\centering
\includegraphics[width=0.65\linewidth]{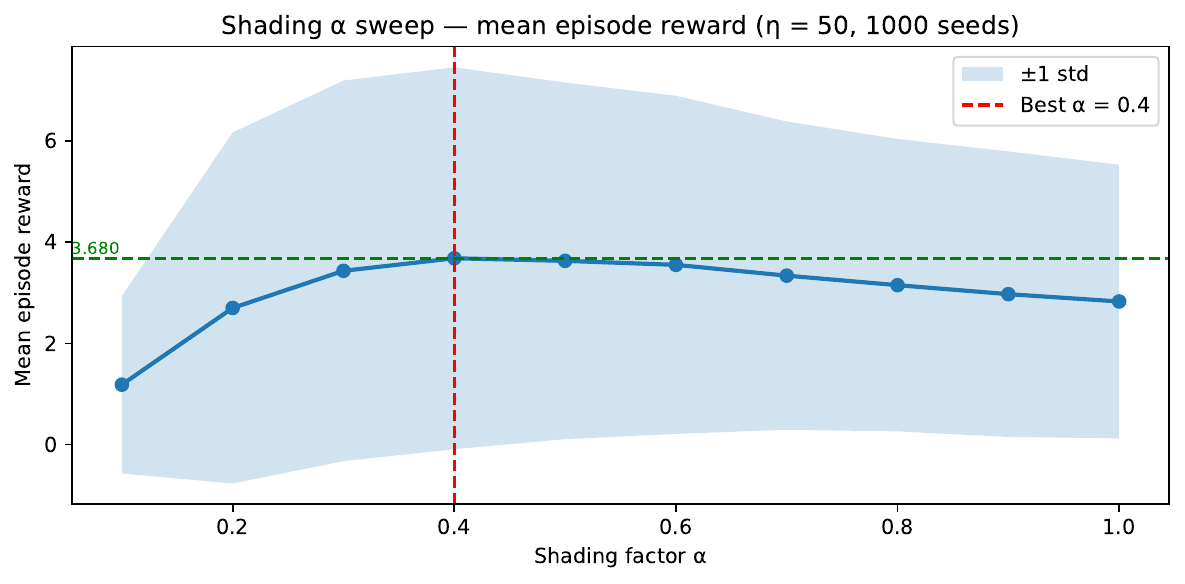}
\caption{Average episode reward of Shading($\alpha$) with $M = 20$ rollouts as a function of the shading multiplier $\alpha$, evaluated over the 1\,000 test campaigns at $\eta = 50$. The shaded band shows $\pm 1$ standard deviation across episodes.}
\label{fig:tune-alpha}
\end{figure}

\paragraph{GreedyMyopic and Shading.} Both policies estimate the value of the current display as the mean difference between $M$ Monte Carlo rollouts that force a win and $M$ that force a loss, continued with the bid 0; each pair of rollouts shares its random numbers, so that environment noise cancels in the difference. The shading multiplier is selected by a grid search over $\alpha \in \{0.1, 0.2, \dots, 1.0\}$ with $M = 20$ rollouts, on the 1\,000 test campaigns: the best value is $\alpha^* = 0.4$ (Figure~\ref{fig:tune-alpha}), and GreedyMyopic is Shading with $\alpha = 1$ and $M = 20$. Unlike the RL hyperparameters, which are selected on validation campaigns, the Shading multiplier (and hence $\alpha_{\rm cont} = 0.4$ in MPC) and the constant bid are selected on the test campaigns, which slightly favors these policies.

\begin{figure}[H]
\centering
\begin{subfigure}[b]{0.49\linewidth}
    \centering
    \includegraphics[width=\linewidth]{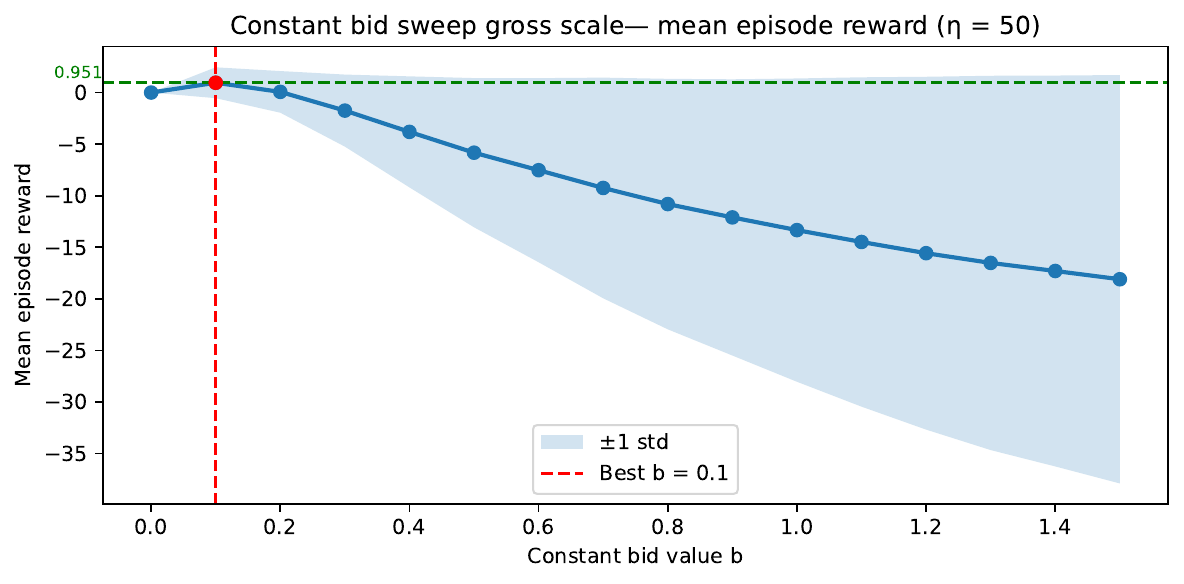}
    \caption{Coarse grid, $b^* = 0.1$, reward = 0.951}
\end{subfigure}
\hfill
\begin{subfigure}[b]{0.49\linewidth}
    \centering
    \includegraphics[width=\linewidth]{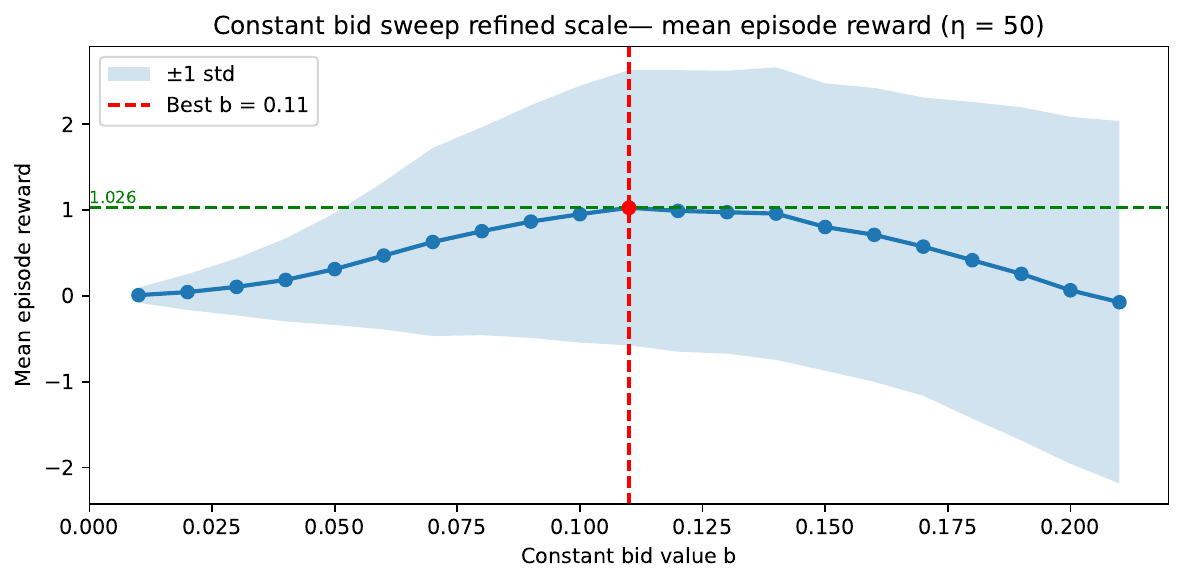}
    \caption{Refined grid, $b^* = 0.11$, reward = 1.026}
\end{subfigure}

\caption{Average episode reward as a function of the constant bid $b$ at $\eta = 50$, using two grid resolutions: coarse ($b$ on a 0.1 grid) and refined (0.01 grid within $\pm 0.1$ of the best coarse value). Results are evaluated over the 1\,000 test campaigns. The shaded band shows $\pm 1$ standard deviation across episodes.}
\label{fig:tune-cst-bid}
\end{figure}

\paragraph{Constant bidder.} The constant bid is selected by a two-stage grid search on the 1\,000 test campaigns, first on a coarse grid of step $0.1$ over $[0, 1.5]$, then on a grid of step $0.01$ within $\pm 0.1$ of the best coarse value: $b^* = 0.11$ (Figure~\ref{fig:tune-cst-bid}).

\subsection{Deep RL training details}
\label{app:rl}

\paragraph{Observations and actions.}
PPO, SAC, and TD3 receive flattened observations from which the oracle value used by \textsc{FiPla} has been removed. This value is unavailable during both training and evaluation. Observations and rewards are normalized using \texttt{VecNormalize}, and bids are clipped to $[0,5]$. We use the default Stable-Baselines3 network architectures: two hidden layers of 64 units with $\tanh$ activations for PPO, and two hidden layers of 256 units with ReLU activations for SAC and TD3.

\paragraph{Hyperparameter search.}
We select the hyperparameters of each algorithm through a grid search at $\eta = 50$, using the Inria Margaret cluster. The search covers 144 configurations for PPO, 48 for SAC, and 96 for TD3. The tested values and fixed parameters are listed in Table~\ref{tab:rl_hyperparams}.

Each configuration is trained for $5\times10^{6}$ timesteps with seed 42 and four parallel environments. During training, we evaluate the policy on 20 campaigns every $10^{4}$ timesteps and save the checkpoint with the highest evaluation reward. At the end of training, we compare this checkpoint with the final checkpoint on a separate set of 200 validation campaigns, using the normalization statistics saved with each checkpoint. For each algorithm, we select the configuration and checkpoint with the highest mean validation reward.

\paragraph{Final training.}
Once the hyperparameters have been selected, we train the chosen configuration again for $10^{7}$ timesteps with three different seeds (42, 43, and 44). For each training, we compare the best intermediate checkpoint and the final checkpoint on the 200 validation campaigns, then evaluate the better one on the 1\,000 test campaigns. The results reported for each RL algorithm are averaged over these three independently trained models.

Table~\ref{tab:rl_hyperparams} reports the selected hyperparameters, their validation rewards during the search, and the test reward obtained by each of the three final trainings.

\begin{table}[H]
\centering
\small
\setlength{\tabcolsep}{6pt}
\caption{Hyperparameter search for PPO, SAC, and TD3 at $\eta=50$. The second column lists the values tested for each parameter or its fixed value. Bold entries indicate the hyperparameters selected using the 200 validation campaigns. The last rows report the validation reward of the selected configuration during the search and the test rewards of its three final trainings.}
\label{tab:rl_hyperparams}
\resizebox{\linewidth}{!}{%
\begin{tabular}{@{}llccc@{}}
\toprule
\textbf{Parameter} & \textbf{Search grid / Fixed} & \textbf{PPO best} & \textbf{SAC best} & \textbf{TD3 best} \\
\midrule
\multicolumn{5}{@{}l}{\textit{Shared}} \\[2pt]
Total timesteps & search $5\times10^{6}$; final $10^{7}$ (3 trainings) & \multicolumn{3}{c}{—} \\
Parallel environments & 4 & \multicolumn{3}{c}{—} \\
Normalization & VecNormalize (observations, rewards) & \multicolumn{3}{c}{—} \\
Eval.\ during training & 20 campaigns every $10^{4}$ steps & \multicolumn{3}{c}{—} \\
Selection & 200 validation campaigns, best or final checkpoint & \multicolumn{3}{c}{—} \\
\midrule
\multicolumn{5}{@{}l}{\textit{PPO}} \\[2pt]
Learning rate & $\{5\!\times\!10^{-5},\,10^{-4},\,2\!\times\!10^{-4},\,3\!\times\!10^{-4}\}$
& $\mathbf{5\times10^{-5}}$ & — & — \\
Rollout length (steps) & $\{256,\,512\}$ & $\mathbf{256}$ & — & — \\
Optimization epochs & $\{2,\,5,\,10\}$ & $\mathbf{10}$ & — & — \\
Entropy coefficient & $\{0,\,0.01,\,0.02\}$ & $\mathbf{0}$ & — & — \\
Discount factor & $\{0.95,\,0.99\}$ & $\mathbf{0.99}$ & — & — \\
Mini-batch size & 128 (fixed) & — & — & — \\
GAE parameter & 0.90 (fixed) & — & — & — \\
Clip range $\epsilon$ & 0.20 (fixed) & — & — & — \\
\midrule
\multicolumn{5}{@{}l}{\textit{SAC}} \\[2pt]
Learning rate & $\{5\!\times\!10^{-5},\,10^{-4},\,3\!\times\!10^{-4},\,10^{-3}\}$
& — & $\mathbf{5\times10^{-5}}$ & — \\
Batch size & $\{256,\,512\}$ & — & $\mathbf{512}$ & — \\
Learning starts (warmup) & $\{500,\,1000,\,1500\}$ & — & $\mathbf{500}$ & — \\
Discount factor & $\{0.95,\,0.99\}$ & — & $\mathbf{0.99}$ & — \\
Entropy coefficient & auto-tuned & — & — & — \\
\midrule
\multicolumn{5}{@{}l}{\textit{TD3}} \\[2pt]
Learning rate & $\{5\!\times\!10^{-5},\,10^{-4},\,3\!\times\!10^{-4},\,10^{-3}\}$
& — & — & $\mathbf{10^{-3}}$ \\
Batch size & $\{256,\,512\}$ & — & — & $\mathbf{512}$ \\
Exploration noise $\sigma$ & $\{0.1,\,0.2\}$ & — & — & $\mathbf{0.2}$ \\
Learning starts (warmup) & $\{500,\,1000,\,1500\}$ & — & — & $\mathbf{500}$ \\
Discount factor & $\{0.95,\,0.99\}$ & — & — & $\mathbf{0.99}$ \\
Policy update delay & 2 (fixed) & — & — & — \\
Target policy noise & 0.2 (fixed) & — & — & — \\
\midrule
\multicolumn{2}{@{}l}{Total grid configurations}
& 144 & 48 & 96 \\
\multicolumn{2}{@{}l}{Validation reward of the selected configuration}
& 4.174 & 0.863 & 3.769 \\
\multicolumn{2}{@{}l}{Test reward of the 3 trainings}
& 3.52 / 3.53 / 1.84 & 0.95 / 0.74 / 0.43 & 2.89 / 3.13 / 0.00 \\
\bottomrule
\end{tabular}}
\end{table}

\begin{figure}[H]
\centering
\includegraphics[width=0.85\linewidth]{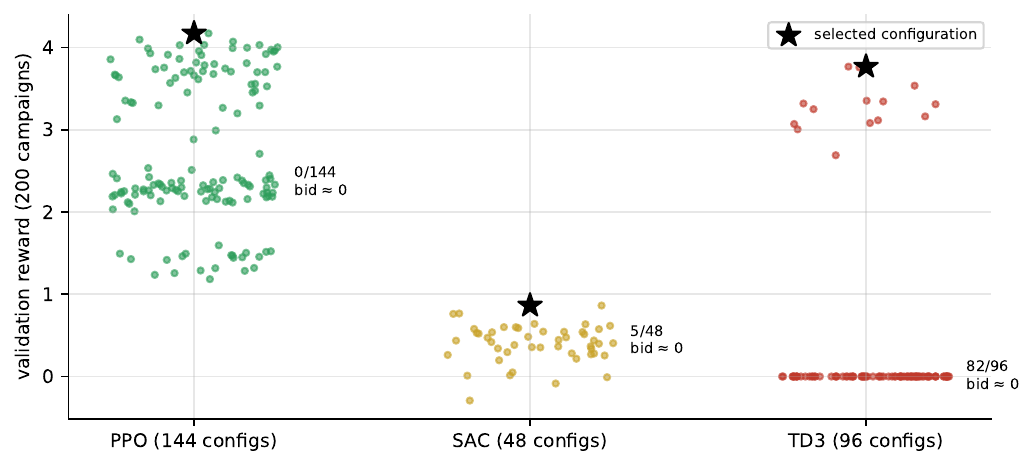}
\caption{Mean reward on 200 validation campaigns for each hyperparameter configuration tested at $\eta=50$. For each configuration, the better of the best intermediate and final checkpoints is shown. Stars indicate the configurations selected for the final trainings. Among the tested configurations, 82 of 96 for TD3 and 5 of 48 for SAC converge to policies that never bid (reward approximately zero).}
\label{fig:rl-search}
\end{figure}

\paragraph{Results of the hyperparameter search.}
Figure~\ref{fig:rl-search} shows the validation reward obtained for each tested configuration. Most TD3 configurations converge to policies that never bid (82 out of 96), as do 5 of the 48 SAC configurations. PPO is less affected by this failure mode, although its final performance still depends on the training seed, as discussed below.

\begin{figure}[H]
\centering
\includegraphics[width=\linewidth]{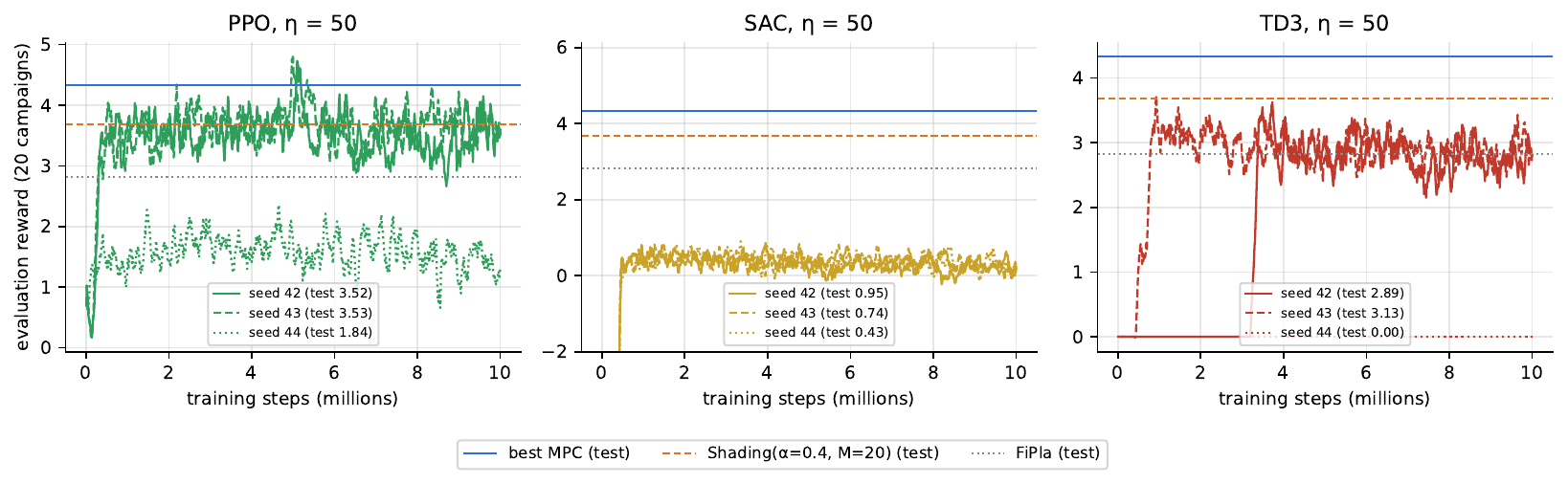}
\caption{Evaluation reward during the three final trainings of PPO, SAC, and TD3 at $\eta=50$. Policies are evaluated on 20 campaigns every $10^{4}$ steps; curves show a rolling mean over 10 evaluations. The legend reports the test reward of each training on 1\,000 campaigns. Horizontal lines indicate the test rewards of the best MPC configuration, Shading($0.4$, $M=20$), and \textsc{FiPla}. The earliest SAC evaluations, with rewards as low as $-50$, are omitted from the vertical scale.}
\label{fig:rl-curves}
\end{figure}

\paragraph{Training behavior.}
Figure~\ref{fig:rl-curves} shows the learning curves of the three final trainings for each algorithm. The runs reveal substantial differences across algorithms and training seeds.

TD3 is particularly unstable: one of its three final trainings converges to a policy that never bids, as observed in most configurations during the search. Once its deterministic bid falls to zero, the agent stops winning auctions and does not recover in these runs.

PPO also shows substantial variability across seeds, with test rewards of $3.52$, $3.53$, and $1.84$ despite using the same hyperparameters. SAC, by contrast, reaches its final performance level within approximately $10^{6}$ steps, but learns a policy that wins many cheap auctions and keeps users fatigued.

\begin{figure}[H]
\centering
\includegraphics[width=0.8\linewidth]{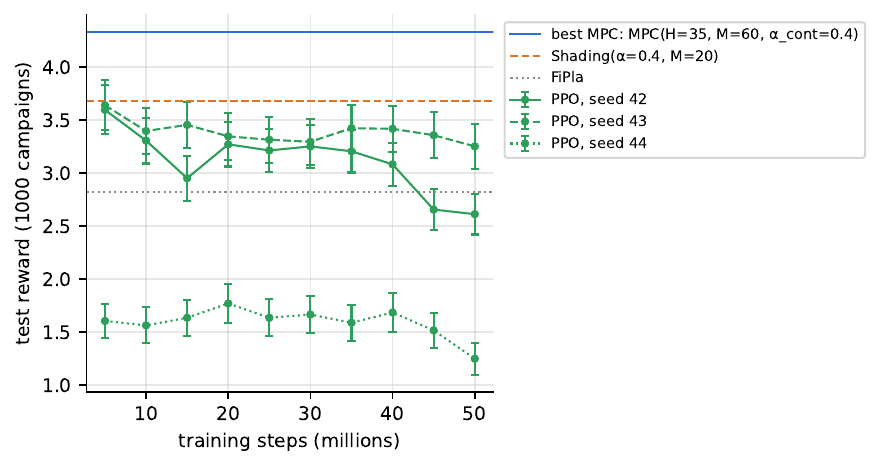}
\caption{Test reward of the selected PPO configuration trained for $5\times10^{7}$ steps at $\eta=50$, using three training seeds. Each point corresponds to a checkpoint saved every $5\times10^{6}$ steps and evaluated on the same 1\,000 test campaigns. Error bars show 95\% confidence intervals.}
\label{fig:ppo-long}
\end{figure}

\paragraph{Longer PPO training.}
To check whether PPO benefits from a larger training budget, we extend the selected configuration to $5\times10^{7}$ timesteps, using the same three training seeds. Figure~\ref{fig:ppo-long} shows that test performance decreases over the extended runs. For example, seed 42 achieves a test reward of $3.59$ after $5\times10^{6}$ steps, $3.31$ after $10^{7}$ steps, and $2.61$ after $5\times10^{7}$ steps. Increasing the training budget therefore does not improve PPO's performance in this experiment.

\section{Additional experiments}
\label{app:additional-experiments}

\subsection{Bidding under varying auction frequency and recovering user fatigue}
\label{sec:burst}

We consider a variant of the main scenario to test whether MPC adjusts its bids when the cost of winning changes over the course of a campaign. Two components are modified :

\begin{itemize}[leftmargin=*, itemsep=1pt, topsep=2pt]
    \item \emph{Recovering fatigue.} Each past win at time $t_i$ reduces the value of future displays through a factor that gradually returns to 1:
    \[
    S_u(t^-) = \prod_{i:\,t_i<t}
    \bigl(1-c e^{-\beta(t-t_i)}\bigr),
    \qquad c=0.9,\quad \beta=0.3.
    \]
    This corresponds to Equation~\eqref{eq:carry-over} with the monoid $([0,1],\times,1)$ and decay map
    $\gamma(x_i,t_i,t)=1-c e^{-\beta(t-t_i)}$.
    Each factor starts at $0.1$ immediately after a win and approaches 1 independently of subsequent wins. After five time units, a win still reduces conversion propensity by about 20\%. The reward kernel uses $S_u(t^-)$ instead of $\phi$, giving a conversion propensity
    $\mathrm{clip}\bigl(S_u(t^-)(\mathbf{1}^\top u)(\mathbf{1}^\top x),0,1\bigr)$.
    The bidder observes $S_u(t^-)$, but not the past win times that determine its future evolution.

    \item \emph{Higher auction frequency at the start of the campaign.}
    Auctions arrive according to a non-homogeneous Poisson process with rate 10 on $[0,10)$ and rate 1 thereafter, instead of rate 1 throughout the campaign. The campaign duration remains exponential with mean $\eta=50$. This produces 131 auctions per campaign on average, including 92 during the initial high-frequency period.
\end{itemize}

During the first ten time units, auctions arrive roughly 0.1 time units apart, so fatigue from successive wins accumulates before it can dissipate. Later, auctions are less frequent and users have more time to recover. Winning an auction can therefore reduce the value of many more future opportunities early in the campaign than later on, making a fixed shading multiplier potentially suboptimal.


\paragraph{Experimental protocol.}
All policies are evaluated on the same 200 test campaigns (environment seeds 0--199, policy seed $s+1000$ for campaign $s$). Comparisons are paired, as in the main benchmark, but use one fifth as many campaigns, giving confidence intervals approximately 2.2 times wider.

Shading uses $M=20$ rollouts, with its multiplier selected from $\{0.1,\ldots,1\}$ on these test campaigns ($\alpha=0.6$). GreedyMyopic corresponds to $\alpha=1$, while the two-stage search for the constant bidder gives $b=0.09$.

MPC uses $M=60$ rollouts per candidate, with $\alpha\in\{0.1,\ldots,1\}$, and a 32-32 ReLU valuation head trained on 1\,000 \textsc{FiPla} campaigns of this scenario, following the main benchmark procedure. We compare three continuation policies:
\begin{itemize}[leftmargin=*, itemsep=1pt, topsep=2pt]
    \item \emph{Fixed:} $\alpha_{\rm cont}=0.6$, the selected Shading multiplier, with $H\in\{5,15,25\}$.
    \item \emph{Shared:} the candidate multiplier $\alpha$ is also used for subsequent bids in each rollout, with $H\in\{5,10,15,25\}$.
    \item \emph{Two-phase:} $\alpha_{\rm cont}=0.3$ before $t=10$ and $0.7$ thereafter, with $H\in\{5,15,25\}$.
\end{itemize}
The two-phase coefficients are selected by searching pairs $(a_1,a_2)$ on a grid with step $0.1$, using an oracle policy that bids $a_1V_0$ before $t=10$ and $a_2V_0$ afterwards. Selection uses 200 validation campaigns (seeds 20\,000--20\,199).

For PPO, we conduct a smaller hyperparameter search centered on the configuration selected in the main benchmark. The initial grid combines learning rates $\{5\times10^{-5},10^{-4},3\times10^{-4}\}$, rollout lengths $\{256,512\}$ and optimization epochs $\{5,10\}$ (12 configurations). Since its best configuration uses 512 steps and 10 epochs, at the edge of this grid, we extend the search to learning rates $\{10^{-4},2\times10^{-4},3\times10^{-4}\}$, rollout lengths $\{512,1024,2048\}$ and epochs $\{5,10,20\}$. This adds 23 new configurations, for 35 in total. The entropy coefficient remains 0 and $\gamma=0.99$; all other settings follow Table~\ref{tab:rl_hyperparams}.

Each configuration is trained for $3\times10^6$ steps and evaluated on 200 validation campaigns using its best intermediate and final checkpoints. The selected configuration (learning rate $10^{-4}$, 1\,024 steps per update, 5 epochs, final checkpoint) is retrained with three seeds for $5\times10^6$ steps, and each training is evaluated on the 200 test campaigns. Both training budgets are shorter than in the main benchmark ($5\times10^6$ and $10^7$ steps, respectively). SAC and TD3 are not evaluated here, as both performed below PPO in the main benchmark.

\begin{table}[H]
\centering
\caption{Burst scenario: mean episode reward ($\pm$ 95\% CI) over the first 200 test campaigns, and paired differences
(RL: mean of the 3 trainings; the test reward of each training in parentheses). MPC continuation (multiplier of the bids
after the first one in each rollout): $\alpha_{\rm cont} = 0.6$ (the best Shading multiplier), shared $\alpha$ (each
rollout bids its candidate $\alpha$ at every auction), or two-phase ($\alpha_{\rm cont} = 0.3$ for $t < 10$, then $0.7$).}
\label{tab:burst}
\footnotesize
\setlength{\tabcolsep}{4pt}
\begin{tabular}{lccc}
\toprule
Policy & reward & $\Delta$ vs Shading($\alpha = 0.6$) & $\Delta$ vs \textsc{FiPla} \\
\midrule
MPC($H = 5$, $M = 60$, $\alpha_{\rm cont} = 0.6$) & $6.531 \pm 0.982$ & $+0.44 \pm 0.26$ & $+1.14 \pm 0.26$ \\
MPC($H = 15$, $M = 60$, $\alpha_{\rm cont} = 0.6$) & $6.471 \pm 0.987$ & $+0.38 \pm 0.27$ & $+1.08 \pm 0.30$ \\
MPC($H = 25$, $M = 60$, $\alpha_{\rm cont} = 0.6$) & $6.453 \pm 0.973$ & $+0.36 \pm 0.27$ & $+1.06 \pm 0.30$ \\
MPC($H = 5$, $M = 60$, shared $\alpha$) & $6.445 \pm 0.974$ & $+0.35 \pm 0.28$ & $+1.05 \pm 0.29$ \\
MPC($H = 10$, $M = 60$, shared $\alpha$) & $6.444 \pm 0.970$ & $+0.35 \pm 0.27$ & $+1.05 \pm 0.30$ \\
MPC($H = 15$, $M = 60$, shared $\alpha$) & $6.354 \pm 0.960$ & $+0.26 \pm 0.26$ & $+0.96 \pm 0.29$ \\
MPC($H = 25$, $M = 60$, shared $\alpha$) & $6.165 \pm 0.943$ & $+0.07 \pm 0.24$ & $+0.77 \pm 0.28$ \\
MPC($H = 5$, $M = 60$, two-phase $\alpha_{\rm cont}$) & $6.456 \pm 0.953$ & $+0.36 \pm 0.28$ & $+1.06 \pm 0.31$ \\
MPC($H = 15$, $M = 60$, two-phase $\alpha_{\rm cont}$) & $6.494 \pm 0.951$ & $+0.40 \pm 0.28$ & $+1.10 \pm 0.31$ \\
MPC($H = 25$, $M = 60$, two-phase $\alpha_{\rm cont}$) & $6.473 \pm 0.953$ & $+0.38 \pm 0.29$ & $+1.08 \pm 0.31$ \\
PPO ($5.80$ / $3.31$ / $5.72$) & $4.947 \pm 0.709$ & $-1.14 \pm 0.39$ & $-0.44 \pm 0.43$ \\
Shading($\alpha = 0.6$, $M = 20$) & $6.091 \pm 0.944$ & --- & $+0.70 \pm 0.28$ \\
\textsc{FiPla} & $5.392 \pm 0.956$ & $-0.70 \pm 0.28$ & --- \\
GreedyMyopic & $5.173 \pm 0.901$ & $-0.92 \pm 0.26$ & $-0.22 \pm 0.25$ \\
Constant($b = 0.09$) & $1.248 \pm 0.242$ & $-4.84 \pm 0.79$ & $-4.14 \pm 0.81$ \\
\bottomrule
\end{tabular}
\end{table}
\paragraph{Results.}
Shading improves on \textsc{FiPla} by $+0.70\pm0.28$ (+13\%). MPC performs better than Shading across the three continuation policies. Its best variant ($H=5$, $M=60$, $\alpha_{\rm cont}=0.6$) gains $+0.44\pm0.26$ (+7\%) over Shading and $+1.14\pm0.26$ (+21\%) over \textsc{FiPla}. The 95\% confidence interval for the gain over Shading excludes zero for 8 of the 10 MPC variants.

MPC bids more selectively during the high-frequency period, winning 5.7\% of auctions, compared with 6.6\% for Shading and 7.7\% for \textsc{FiPla}. After $t=10$, its win rate rises to 17.9\%, close to Shading's 17.6\%. With the shared continuation, MPC selects an average multiplier of 0.45 before $t=10$ and 0.74 afterwards, close to the oracle two-phase coefficients $(0.3,0.7)$.

However, allowing the continuation policy to vary with time brings no clear improvement over the fixed multiplier $\alpha_{\rm cont}=0.6$. Most variants perform similarly, and the shared continuation at $H=25$ is the only one significantly below the best fixed-continuation variant ($-0.37\pm0.23$). Replanning at every auction appears sufficient to adapt the current bid to the campaign phase, without requiring a time-dependent continuation policy or a longer planning horizon.

PPO remains below Shading on average ($-1.14\pm0.39$) and below the best MPC ($-1.58\pm0.41$). Two of its three trainings outperform \textsc{FiPla} ($+0.41\pm0.33$ and $+0.33\pm0.38$) but remain below Shading ($-0.29\pm0.29$ and $-0.37\pm0.36$). The third learns to bid too little, winning only 5.2\% of auctions, compared with 9.3--12.2\% for the other policies except Constant.

The selected PPO run reaches a validation reward of 6.62 after $3\times10^6$ steps, but its three retrainings achieve only 6.10, 2.46 and 6.29 on the same validation campaigns after $5\times10^6$ steps. The high reward of the selected run thus appears optimistic, reflecting the selection of the best among 35 configurations on only 200 validation campaigns.

\subsection{Computational cost and performance of bidding policies}
Table~\ref{tab:timing} reports the time to compute one bid, measured on the same 80 decision states of the main scenario
($\eta = 50$), on one CPU core of a laptop (Intel Core Ultra 7 265H) and of a cluster node (Intel Xeon E5-2695 v3); the
time per campaign is the time per bid times 47.7 auctions. MPC's cost is linear in the number of rollouts $M$ and in the
horizon $H$ (each rollout simulates the next $H$ time units), whereas its reward saturates around $H = 25$ and
$M = 40$--$80$; the best setting, MPC($H = 35$, $M = 60$, $\alpha_{\rm cont} = 0.4$), costs about 2.3\,s per bid on the
laptop. Changing $\alpha_{\rm cont}$ does not change the cost. One prediction of the valuation head takes $0.13$\,ms, and
the RL policies need less than 1\,ms per bid.

\begin{table}[H]
\centering
\caption{Decision time per bid (mean over 80 decision states, one CPU core) and mean episode reward at $\eta = 50$.}
\label{tab:timing}
\footnotesize
\setlength{\tabcolsep}{4pt}
\resizebox{\linewidth}{!}{%
\begin{tabular}{lcccc}
\toprule
Policy & ms / bid (laptop) & ms / bid (cluster node) & s / campaign (laptop) & reward ($\eta = 50$) \\
\midrule
FiPla & $<0.01$ & $<0.01$ & $<0.01$ & $2.821 \pm 0.160$ \\
Shading($\alpha = 0.4$, $M = 20$) & 66.1 & 147 & 3.15 & $3.680 \pm 0.234$ \\
Shading($\alpha = 0.4$, $M = 40$) & 133 & 297 & 6.33 & $3.680 \pm 0.234$ \\
Shading($\alpha = 0.4$, $M = 60$) & 202 & 446 & 9.64 & $3.654 \pm 0.232$ \\
MPC($H = 1$, $M = 20$) & 155 & 308 & 7.40 & $3.458 \pm 0.207$ \\
MPC($H = 1$, $M = 40$) & 311 & 611 & 14.8 & $3.360 \pm 0.193$ \\
MPC($H = 1$, $M = 100$) & 769 & 1\,538 & 36.7 & $3.309 \pm 0.190$ \\
MPC($H = 1$, $M = 200$) & 1\,570 & 3\,067 & 74.9 & $3.307 \pm 0.190$ \\
MPC($H = 10$, $M = 20$) & 373 & 878 & 17.8 & $3.911 \pm 0.236$ \\
MPC($H = 10$, $M = 20$, $\alpha_{\rm cont} = 0.4$) & \multicolumn{3}{c}{same cost as $\alpha_{\rm cont} = 1$} & $3.898 \pm 0.239$ \\
MPC($H = 10$, $M = 40$) & 734 & 1\,745 & 35.0 & $3.862 \pm 0.226$ \\
MPC($H = 10$, $M = 40$, $\alpha_{\rm cont} = 0.4$) & \multicolumn{3}{c}{same cost as $\alpha_{\rm cont} = 1$} & $3.829 \pm 0.228$ \\
MPC($H = 10$, $M = 60$) & 1\,120 & 2\,615 & 53.4 & $3.938 \pm 0.238$ \\
MPC($H = 25$, $M = 20$) & 639 & 1\,587 & 30.5 & $4.040 \pm 0.256$ \\
MPC($H = 25$, $M = 20$, $\alpha_{\rm cont} = 0.4$) & \multicolumn{3}{c}{same cost as $\alpha_{\rm cont} = 1$} & $4.120 \pm 0.264$ \\
MPC($H = 25$, $M = 40$) & 1\,294 & 3\,156 & 61.7 & $4.153 \pm 0.262$ \\
MPC($H = 25$, $M = 40$, $\alpha_{\rm cont} = 0.4$) & \multicolumn{3}{c}{same cost as $\alpha_{\rm cont} = 1$} & $4.228 \pm 0.263$ \\
MPC($H = 25$, $M = 60$) & 1\,932 & 4\,742 & 92.2 & $4.112 \pm 0.257$ \\
MPC($H = 25$, $M = 60$, $\alpha_{\rm cont} = 0.4$) & \multicolumn{3}{c}{same cost as $\alpha_{\rm cont} = 1$} & $4.285 \pm 0.268$ \\
MPC($H = 25$, $M = 80$) & 2\,523 & 6\,320 & 120 & $4.282 \pm 0.268$ \\
MPC($H = 35$, $M = 20$) & 781 & 1\,950 & 37.2 & $4.030 \pm 0.260$ \\
MPC($H = 35$, $M = 40$) & 1\,539 & 3\,885 & 73.4 & $4.225 \pm 0.273$ \\
MPC($H = 35$, $M = 60$) & 2\,278 & 5\,859 & 109 & $4.206 \pm 0.269$ \\
MPC($H = 35$, $M = 60$, $\alpha_{\rm cont} = 0.4$) & \multicolumn{3}{c}{same cost as $\alpha_{\rm cont} = 1$} & $4.328 \pm 0.275$ \\
PPO (inference) & 0.574 & 0.739 & 0.0274 & $2.965 \pm 0.182$ \\
SAC (inference) & 0.441 & 0.697 & 0.0211 & $0.707 \pm 0.119$ \\
TD3 (inference) & 0.302 & 0.538 & 0.0144 & $2.006 \pm 0.121$ \\
\bottomrule
\end{tabular}}
\end{table}

\subsection{Robustness of MPC to valuation errors and fewer rollouts}

We study how errors in the learned valuation head $\widehat V_0$ and the number of planning rollouts affect MPC's performance. Our reference is MPC($H=25$, $M=80$, $\alpha_{\rm cont}=1$), using the 32-32 ReLU head trained on 1\,000 \textsc{FiPla} campaigns. We consider four types of changes (Table~\ref{tab:v0}): (i) multiplicative noise, scaling each prediction by $e^{\sigma\varepsilon-\sigma^2/2}$, with $\varepsilon\sim\mathcal{N}(0,1)$ and $\sigma\in\{0.1,0.25,0.5,1\}$; (ii) heads trained on only 10, 30, 100, or 300 campaigns; (iii) misspecified heads that ignore fatigue ($S=1$), are trained with higher competition ($\mathrm{LogNormal}(0,1)$), or have their predictions multiplied by 0.5 or 2; and (iv) fewer planning rollouts, keeping the reference head unchanged.

All variants are evaluated on the same 1\,000 test campaigns, with paired differences and 95\% confidence intervals. The valuation-head experiments also use the same rollout random numbers as the reference. Prediction errors are measured against the oracle $V_0$ on auctions with $V_0>0.05$.

\begin{table}[H]
\centering
\caption{Performance of MPC under valuation errors and reduced rollout counts. The reference uses $H=25$, $M=80$ and the 32-32 ReLU valuation head trained on 1\,000 campaigns. Correlation and mean absolute error are measured against the oracle $V_0$.}
\label{tab:v0}
\footnotesize
\setlength{\tabcolsep}{3.5pt}
\resizebox{\linewidth}{!}{%
\begin{tabular}{llcccccc}
\toprule
family & variant & corr. & $|\widehat V_0 - V_0|$ & reward & $\Delta$ vs reference & $\Delta$ vs Shading \\
\midrule
reference & exact head (1\,000 campaigns), $M = 80$ & 0.97 & 0.050 & $4.282 \pm 0.268$ & --- & $+0.60 \pm 0.13$ \\
noisy $\widehat V_0$ & $\sigma = 0.1$ & 0.95 & 0.069 & $4.117 \pm 0.260$ & $-0.17 \pm 0.10$ & $+0.44 \pm 0.12$ \\
 & $\sigma = 0.25$ & 0.88 & 0.125 & $4.118 \pm 0.261$ & $-0.16 \pm 0.10$ & $+0.44 \pm 0.13$ \\
 & $\sigma = 0.5$ & 0.69 & 0.225 & $3.930 \pm 0.251$ & $-0.35 \pm 0.10$ & $+0.25 \pm 0.11$ \\
 & $\sigma = 1$ & 0.34 & 0.455 & $3.579 \pm 0.244$ & $-0.70 \pm 0.11$ & $-0.10 \pm 0.12$ \\
fewer training campaigns & 10 campaigns & 0.16 & 0.470 & $2.179 \pm 0.218$ & $-2.10 \pm 0.22$ & $-1.50 \pm 0.21$ \\
 & 30 campaigns & 0.62 & 0.265 & $3.774 \pm 0.238$ & $-0.51 \pm 0.11$ & $+0.09 \pm 0.12$ \\
 & 100 campaigns & 0.75 & 0.212 & $4.195 \pm 0.262$ & $-0.09 \pm 0.09$ & $+0.52 \pm 0.11$ \\
 & 300 campaigns & 0.94 & 0.082 & $4.250 \pm 0.268$ & $-0.03 \pm 0.09$ & $+0.57 \pm 0.13$ \\
misspecified head & ignores fatigue ($S = 1$) & 0.76 & 0.167 & $3.799 \pm 0.261$ & $-0.48 \pm 0.11$ & $+0.12 \pm 0.14$ \\
 & trained on higher competition & 0.98 & 0.039 & $4.233 \pm 0.267$ & $-0.05 \pm 0.08$ & $+0.55 \pm 0.12$ \\
scaled $\widehat V_0$ & $\widehat V_0 \times 0.5$ & 0.96 & 0.314 & $4.350 \pm 0.293$ & $+0.07 \pm 0.10$ & $+0.67 \pm 0.14$ \\
 & $\widehat V_0 \times 2$ & 0.97 & 0.492 & $3.827 \pm 0.234$ & $-0.45 \pm 0.12$ & $+0.15 \pm 0.12$ \\
planning rollouts & $M = 10$ & --- & --- & $3.881 \pm 0.252$ & $-0.40 \pm 0.12$ & $+0.20 \pm 0.11$ \\
 & $M = 20$ & --- & --- & $4.040 \pm 0.256$ & $-0.24 \pm 0.10$ & $+0.36 \pm 0.12$ \\
 & $M = 40$ & --- & --- & $4.153 \pm 0.262$ & $-0.13 \pm 0.10$ & $+0.47 \pm 0.13$ \\
 & $M = 60$ & --- & --- & $4.112 \pm 0.257$ & $-0.17 \pm 0.11$ & $+0.43 \pm 0.11$ \\
baseline & Shading($\alpha = 0.4$, $M = 20$) & --- & --- & $3.680 \pm 0.234$ & $-0.60 \pm 0.13$ & --- \\
\bottomrule
\end{tabular}}
\end{table}

\begin{figure}[H]
\centering
\includegraphics[width=0.75\linewidth]{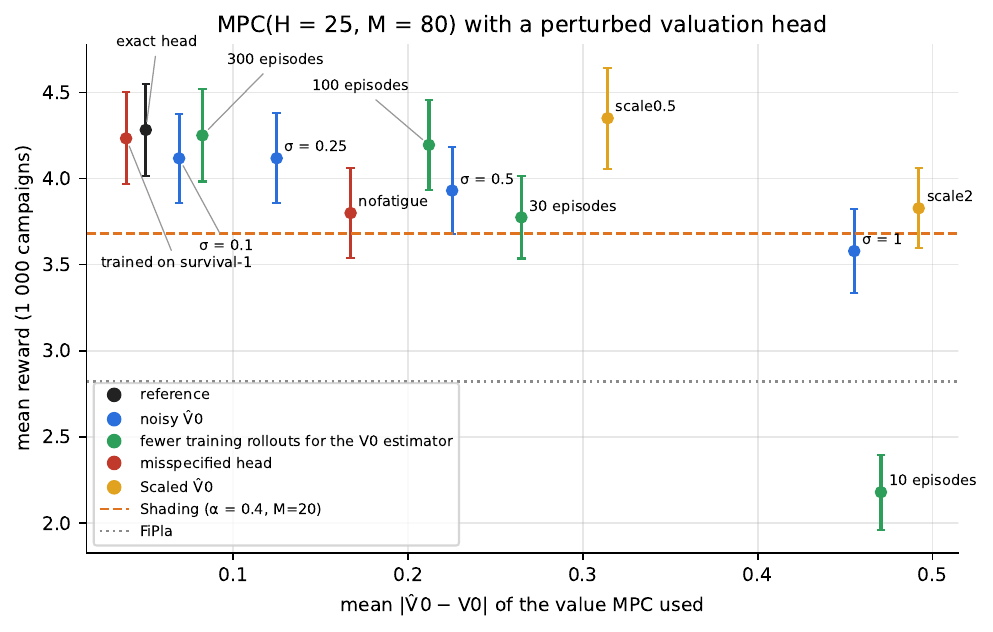}
\caption{Mean episode reward of MPC($H=25$, $M=80$) as a function of the mean absolute error of its valuation head $\widehat V_0$, for the different perturbations.}
\label{fig:v0}
\end{figure}

MPC tolerates moderate valuation errors (Table~\ref{tab:v0}, Figure~\ref{fig:v0}). Heads trained on 100 or 300 campaigns perform close to the reference ($-0.09\pm0.09$ and $-0.03\pm0.09$). Multiplicative noise reduces the reward by about 0.17 for $\sigma\leq0.25$ and 0.35 for $\sigma=0.5$, but MPC still outperforms Shading. With $\sigma=1$, it falls to Shading's level; with only 10 training campaigns, it performs substantially worse. Ignoring fatigue reduces the reward by $0.48\pm0.11$, bringing MPC close to Shading, whereas a head trained under higher competition performs similarly to the reference ($-0.05\pm0.08$).

Errors in the scale of $\widehat V_0$ have asymmetric effects. Halving its predictions does not reduce performance ($+0.07\pm0.10$), as it gives a finer grid of candidate bids in the relevant range. Doubling them leads to excessive bids, including during rollout continuation, and reduces the reward by $0.45\pm0.12$. Reducing the number of rollouts has a smaller effect than large valuation errors: using $M=20$--$60$ instead of 80 lowers the reward by 0.13--0.24, while even $M=10$ remains above Shading ($+0.20\pm0.11$).

\subsection{Evaluating bidding policies without delayed rewards}

We evaluate a variant of the main scenario in which conversions occur immediately after a winning auction, with the same probability $f$ as in the original scenario (Table~\ref{tab:immediate-scenario}). This removes delayed credit assignment while keeping the fatigue dynamics, competition, and campaign lengths unchanged. The variant, \texttt{survival-second-price-exponential-immediate}, also ensures that no conversion is lost after the last auction. MPC uses a valuation head trained on this scenario, while the RL agents follow the same hyperparameter search and training protocol as in the main benchmark.

\begin{table}[H]
\centering
\caption{Comparison of the main scenario and its immediate-conversion variant. Only the timing and reporting of conversions change; the fatigue dynamics and conversion probabilities remain unchanged.}
\label{tab:immediate-scenario}
\footnotesize
\setlength{\tabcolsep}{4pt}
\renewcommand{\arraystretch}{1.15}

\begin{tabularx}{\linewidth}{
@{}
>{\raggedright\arraybackslash}p{0.28\linewidth}
>{\raggedright\arraybackslash}X
>{\raggedright\arraybackslash}p{0.23\linewidth}
@{}
}
\toprule
 & Main scenario & Immediate rewards \\
\midrule
Carry-over context
& $S_u(t^-) = \{t_i < t\}$, the past wins
& Same \\

Conversion probability of a won display
& $\begin{aligned}
f &= \operatorname{clip}\bigl(\\
  &\quad \phi(S_u(t^-),t)\\
  &\quad \cdot(\mathbf{1}^\top u)(\mathbf{1}^\top x),\,0,1\bigr)
\end{aligned}$
& Same $f$ \\

Conversions per won display
& At most 1
& At most 1 \\

Time of conversion
& $t+s$, delay $s\sim\mathrm{Exp}(1)$
& $t$ (no delay) \\

Conversions after the last auction
& Never credited (including after $T$)
& Cannot occur \\

Reward received by the agent
& Payment at the winning auction; conversion value at the first auction after $t+s$
& Both at the winning auction \\

Oracle value $V_0$
& $f(1-e^{-(T-t)})$
& $f$ \\
\bottomrule
\end{tabularx}
\end{table}

\begin{table}[H]
\centering
\caption{Performance of the bidding policies with immediate conversions. Results are mean episode rewards ($\pm$ 95\% CI) over 1\,000 test campaigns, with paired differences relative to MPC and Shading. For RL, the mean is taken over three trainings, whose individual test rewards are shown in parentheses.}
\label{tab:immediate}
\footnotesize
\begin{tabular}{lccc}
\toprule
Policy & reward & $\Delta$ vs MPC & $\Delta$ vs Shading \\
\midrule
MPC($H = 25$, $M = 80$) & $4.409 \pm 0.258$ & --- & $+0.59 \pm 0.13$ \\
PPO (4.11 / 4.23 / 3.86) & $4.069 \pm 0.223$ & $-0.34 \pm 0.10$ & $+0.25 \pm 0.10$ \\
Shading(best $\alpha = 0.4$, $M = 20$) & $3.822 \pm 0.223$ & $-0.59 \pm 0.13$ & --- \\
FiPla & $2.961 \pm 0.161$ & $-1.45 \pm 0.18$ & $-0.86 \pm 0.15$ \\
GreedyMyopic & $2.908 \pm 0.155$ & $-1.50 \pm 0.18$ & $-0.91 \pm 0.15$ \\
TD3 (3.93 / 0.00 / 0.00) & $1.309 \pm 0.072$ & $-3.10 \pm 0.20$ & $-2.51 \pm 0.17$ \\
Constant(best $b = 0.11$) & $1.092 \pm 0.103$ & $-3.32 \pm 0.21$ & $-2.73 \pm 0.19$ \\
\bottomrule
\end{tabular}
\end{table}

With immediate conversions, PPO outperforms all myopic baselines, including Shading ($+0.25 \pm 0.10$), but remains below MPC ($-0.34 \pm 0.10$). Thus, removing delayed feedback does not eliminate the advantage of MPC, which can still plan around the effects of fatigue on future auctions. TD3 remains unstable: one training reaches 3.93, above Shading, while the other two converge to policies that never bid, as did 44 of its 96 hyperparameter configurations during the search.

Immediate conversions also prevent rewards from being lost after the last auction, unlike in the main scenario. Since the two scenarios do not share the same random draws, performance comparisons are made within each scenario.

\subsection{Evaluating bidding policies against the optimal policy under saturation and recency}

We consider two simpler scenarios in which the optimal bidding policy can be computed by dynamic programming (DP). Neither scenario has user or display features, and the carry-over state is one-dimensional. They differ in how past wins affect conversion probability:
\begin{itemize}[leftmargin=*, itemsep=1pt, topsep=2pt]
    \item \emph{Saturation:} $p(n)=e^{-n/10}$, where $n$ is the number of past wins.
    \item \emph{Recency:} $p(\tau)=1-e^{-0.5\tau}$, where $\tau$ is the time since the last win.
\end{itemize}
Both scenarios use exponentially distributed conversion delays with rate $0.5$ and competing bids drawn from $\mathrm{LogNormal}(-1,1)$.

We compare the bidding policies with the DP optimum on the same 1\,000 test campaigns. Unless otherwise stated, MPC uses $M=60$ rollouts and $\alpha_{\rm cont}=1$, while the RL agents follow the training protocol of the main benchmark. We also evaluate PPO with the time since the last win added to its observations, and a diagnostic Cooldown policy that bids the oracle $V_0$ except during a fixed period after each win. The tables report the fraction of the improvement over \textsc{FiPla} achieved by each policy, defined as $(R-R_{\mathrm{FiPla}})/(R_{\mathrm{DP}}-R_{\mathrm{FiPla}})$, where $R$ denotes the mean episode reward.

\paragraph{Computing the optimal policy.}
In both scenarios, the campaign duration is exponential with mean $\eta=50$, giving a termination rate $h=1/\eta=0.02$. Auction arrivals have rate 1. Because the remaining campaign duration is memoryless, the optimal policy depends only on the current state $s$: the number of past wins $n$ for saturation, or the time since the last win $\tau$ for recency. Before the first win, $\tau=\infty$ and $p(\tau)=1$.

A display won in state $s$ has expected conversion value
\[
r(s)=p(s)\frac{\delta}{\delta+h},
\]
where $\delta=0.5$ is the conversion-delay rate. The factor $\delta/(\delta+h)$ accounts for the probability that the conversion occurs before the campaign ends.

The probability of reaching another auction before the campaign ends is $q=1/(1+h)$. Conditional on reaching it, the elapsed time is $G\sim\mathrm{Exp}(1+h)$. A win changes the state to $s'_{\rm win}=n+1$ in saturation or $s'_{\rm win}=G$ in recency; after a loss, the state is $s'_{\rm lose}=n$ or $s'_{\rm lose}=\tau+G$, respectively.

In a second-price auction, the optimal bid is the marginal value of winning:
\[
b^*(s)=\max(D(s),0), \qquad
D(s)=r(s)+q\bigl(\mathbb{E}V(s'_{\rm win})
                    -\mathbb{E}V(s'_{\rm lose})\bigr),
\]
where $V(s)$ is the optimal expected future reward and satisfies
\[
V(s)=q\,\mathbb{E}V(s'_{\rm lose})
       +\mathbb{E}\bigl[(D(s)-B)^+\bigr],
\qquad B\sim\mathrm{LogNormal}(-1,1).
\]
The last expectation has a closed form. For saturation, we solve the Bellman equation by backward induction from $n=200$, where conversion probability is negligible, using bisection to solve for each $V(n)$. For recency, we use value iteration on a grid $\tau\in[0,60]$ with step $0.02$, stopping when the maximum change falls below $10^{-9}$.

The resulting policy bids $b^*(s)$ at every auction. Since it uses the known model, it provides an optimal reference in expectation, subject to numerical approximation and test-sample variability. Table~\ref{tab:dp-alpha} reports its effective shading multiplier $\alpha^*(s)=b^*(s)/r(s)$.

\paragraph{Why constant shading is close to optimal.}
In both scenarios, the optimal bid is approximately a fixed fraction of the myopic value $r(s)$ (Table~\ref{tab:dp-alpha}). The effective multiplier $\alpha^*(s)=b^*(s)/r(s)$ ranges from 0.39 to 0.66 for the first 15 wins in saturation, and from 0.53 to 0.64 for the recency states considered. A single shading multiplier therefore approximates the optimal policy well. Indeed, Shading is statistically indistinguishable from the DP optimum in both scenarios: $+0.03\pm0.08$ in saturation and $-0.09\pm0.12$ in recency.

With $M=60$ rollouts, MPC remains below Shading in both scenarios. Increasing $M$ to 1\,000 substantially reduces this gap, particularly in recency, where MPC matches Shading ($+0.00\pm0.12$). This suggests that the accuracy of MPC's rollout estimates matters when the optimal multiplier varies little across states.

\begin{table}[H]
\centering
\caption{Saturation kernel (RL: mean of the 3 trainings, the test reward of each in parentheses). MPC: $\alpha_{\rm cont} = 1$ unless otherwise stated.}
\label{tab:saturation}
\footnotesize
\setlength{\tabcolsep}{3.5pt}
\begin{tabular}{lcccc}
\toprule
Policy & reward & $\Delta$ vs DP optimum & $\Delta$ vs Shading & share of DP gain \\
\midrule
Shading(best $\alpha = 0.5$, $M = 20$) & $4.351 \pm 0.158$ & $+0.03 \pm 0.08$ & --- & 105\% \\
DP optimum & $4.323 \pm 0.162$ & --- & $-0.03 \pm 0.08$ & 100\% \\
MPC($H = 25$, $M = 1\,000$) & $4.237 \pm 0.157$ & $-0.09 \pm 0.09$ & $-0.11 \pm 0.10$ & 86\% \\
PPO (4.20 / 4.08 / 4.24) & $4.173 \pm 0.143$ & $-0.15 \pm 0.07$ & $-0.18 \pm 0.08$ & 75\% \\
MPC($H = 25$, $M = 200$) & $4.138 \pm 0.150$ & $-0.18 \pm 0.10$ & $-0.21 \pm 0.10$ & 69\% \\
MPC($H = 25$, $M = 80$) & $4.097 \pm 0.154$ & $-0.23 \pm 0.11$ & $-0.25 \pm 0.10$ & 62\% \\
MPC($H = 10$, $M = 60$, $\alpha_{\rm cont} = 0.5$) & $4.097 \pm 0.152$ & $-0.23 \pm 0.10$ & $-0.25 \pm 0.11$ & 62\% \\
MPC($H = 10$, $M = 60$) & $4.056 \pm 0.154$ & $-0.27 \pm 0.11$ & $-0.30 \pm 0.10$ & 56\% \\
MPC($H = 35$, $M = 60$) & $4.051 \pm 0.156$ & $-0.27 \pm 0.10$ & $-0.30 \pm 0.10$ & 55\% \\
MPC($H = 25$, $M = 60$, $\alpha_{\rm cont} = 0.5$) & $4.040 \pm 0.148$ & $-0.28 \pm 0.10$ & $-0.31 \pm 0.10$ & 53\% \\
MPC($H = 35$, $M = 60$, $\alpha_{\rm cont} = 0.5$) & $4.039 \pm 0.149$ & $-0.28 \pm 0.10$ & $-0.31 \pm 0.10$ & 53\% \\
MPC($H = 25$, $M = 60$) & $4.029 \pm 0.154$ & $-0.29 \pm 0.11$ & $-0.32 \pm 0.11$ & 51\% \\
Constant(best $b = 0.18$) & $3.856 \pm 0.169$ & $-0.47 \pm 0.12$ & $-0.50 \pm 0.12$ & 22\% \\
FiPla & $3.721 \pm 0.142$ & $-0.60 \pm 0.12$ & $-0.63 \pm 0.12$ & 0\% \\
GreedyMyopic & $3.605 \pm 0.151$ & $-0.72 \pm 0.13$ & $-0.75 \pm 0.12$ & -19\% \\
\bottomrule
\end{tabular}
\end{table}

\begin{table}[H]
\centering
\caption{Recency kernel (RL: mean of the 3 trainings, the test reward of each in parentheses). MPC: $\alpha_{\rm cont} = 1$ unless otherwise stated.}
\label{tab:recency}
\footnotesize
\setlength{\tabcolsep}{3.5pt}
\begin{tabular}{lcccc}
\toprule
Policy & reward & $\Delta$ vs DP optimum & $\Delta$ vs Shading & share of DP gain \\
\midrule
DP optimum & $7.875 \pm 0.529$ & --- & $+0.09 \pm 0.12$ & 100\% \\
MPC($H = 25$, $M = 1\,000$) & $7.786 \pm 0.521$ & $-0.09 \pm 0.12$ & $+0.00 \pm 0.12$ & 89\% \\
Shading(best $\alpha = 0.6$, $M = 20$) & $7.784 \pm 0.520$ & $-0.09 \pm 0.12$ & --- & 88\% \\
MPC($H = 25$, $M = 200$) & $7.670 \pm 0.516$ & $-0.20 \pm 0.13$ & $-0.11 \pm 0.13$ & 74\% \\
PPO + time since last win (7.66 / 7.63 / 7.63) & $7.643 \pm 0.510$ & $-0.23 \pm 0.10$ & $-0.14 \pm 0.10$ & 70\% \\
MPC($H = 25$, $M = 80$) & $7.542 \pm 0.510$ & $-0.33 \pm 0.13$ & $-0.24 \pm 0.14$ & 57\% \\
MPC($H = 10$, $M = 60$, $\alpha_{\rm cont} = 0.6$) & $7.498 \pm 0.505$ & $-0.38 \pm 0.13$ & $-0.29 \pm 0.13$ & 52\% \\
MPC($H = 25$, $M = 60$, $\alpha_{\rm cont} = 0.6$) & $7.488 \pm 0.504$ & $-0.39 \pm 0.13$ & $-0.30 \pm 0.13$ & 50\% \\
MPC($H = 35$, $M = 60$, $\alpha_{\rm cont} = 0.6$) & $7.488 \pm 0.504$ & $-0.39 \pm 0.13$ & $-0.30 \pm 0.13$ & 50\% \\
MPC($H = 25$, $M = 60$) & $7.483 \pm 0.505$ & $-0.39 \pm 0.14$ & $-0.30 \pm 0.13$ & 50\% \\
MPC($H = 35$, $M = 60$) & $7.483 \pm 0.505$ & $-0.39 \pm 0.14$ & $-0.30 \pm 0.13$ & 50\% \\
MPC($H = 10$, $M = 60$) & $7.476 \pm 0.505$ & $-0.40 \pm 0.14$ & $-0.31 \pm 0.13$ & 49\% \\
Constant(best $b = 0.31$) & $7.099 \pm 0.489$ & $-0.78 \pm 0.15$ & $-0.68 \pm 0.15$ & 1\% \\
FiPla & $7.095 \pm 0.490$ & $-0.78 \pm 0.18$ & $-0.69 \pm 0.17$ & 0\% \\
PPO (7.09 / 7.02 / 7.03) & $7.048 \pm 0.464$ & $-0.83 \pm 0.14$ & $-0.74 \pm 0.13$ & -6\% \\
Cooldown(oracle $V_0$, best $\tau = 1$) & $6.998 \pm 0.473$ & $-0.88 \pm 0.17$ & $-0.79 \pm 0.15$ & -12\% \\
GreedyMyopic & $6.856 \pm 0.478$ & $-1.02 \pm 0.18$ & $-0.93 \pm 0.16$ & -31\% \\
\bottomrule
\end{tabular}
\end{table}

\begin{table}[H]
\centering
\caption{Optimal effective shading $\alpha^* = b^*/r$ of the DP by state ($n$: past wins; $\tau$: time since the last win).}
\label{tab:dp-alpha}
\footnotesize
\begin{tabular}{lcccccc}
\toprule
saturation, $n$ & 0 & 2 & 5 & 10 & 15 & 20 \\
\quad $\alpha^* = b^*/r$ & 0.39 & 0.42 & 0.46 & 0.55 & 0.66 & 0.76 \\
recency, $\tau$ & 0.25 & 1 & 2 & 3 & 5 & 10 \\
\quad $\alpha^* = b^*/r$ & 0.64 & 0.60 & 0.57 & 0.55 & 0.54 & 0.53 \\
\bottomrule
\end{tabular}
\end{table}

\paragraph{Effect of state representation on PPO.}
PPO performs close to Shading in saturation, where the number of past wins is directly available in its observation. In recency, however, PPO receives the current time and the time of the last win as separate inputs (with $-1000$ indicating no previous win). Its bids barely depend on their difference, the time elapsed since the last win, and its performance is close to that of a constant bidder. Adding this elapsed time explicitly to the observation, while keeping the same hyperparameters and training seeds, improves PPO's reward by about 0.6 and brings it close to Shading ($-0.14\pm0.10$). The gap in this scenario is therefore largely explained by the state representation.

\section*{NeurIPS Paper Checklist}

\begin{enumerate}

\item {\bf Claims}
    \item[] Question: Do the main claims made in the abstract and introduction accurately reflect the paper's contributions and scope?
    \item[] Answer: \answerYes{} 
    \item[] Justification: it does 
    \item[] Guidelines:
    \begin{itemize}
        \item The answer \answerNA{} means that the abstract and introduction do not include the claims made in the paper.
        \item The abstract and/or introduction should clearly state the claims made, including the contributions made in the paper and important assumptions and limitations. A \answerNo{} or \answerNA{} answer to this question will not be perceived well by the reviewers. 
        \item The claims made should match theoretical and experimental results, and reflect how much the results can be expected to generalize to other settings. 
        \item It is fine to include aspirational goals as motivation as long as it is clear that these goals are not attained by the paper. 
    \end{itemize}

\item {\bf Limitations}
    \item[] Question: Does the paper discuss the limitations of the work performed by the authors?
    \item[] Answer: \answerYes{} 
    \item[] Justification: it does 
    \item[] Guidelines:
    \begin{itemize}
        \item The answer \answerNA{} means that the paper has no limitation while the answer \answerNo{} means that the paper has limitations, but those are not discussed in the paper. 
        \item The authors are encouraged to create a separate ``Limitations'' section in their paper.
        \item The paper should point out any strong assumptions and how robust the results are to violations of these assumptions (e.g., independence assumptions, noiseless settings, model well-specification, asymptotic approximations only holding locally). The authors should reflect on how these assumptions might be violated in practice and what the implications would be.
        \item The authors should reflect on the scope of the claims made, e.g., if the approach was only tested on a few datasets or with a few runs. In general, empirical results often depend on implicit assumptions, which should be articulated.
        \item The authors should reflect on the factors that influence the performance of the approach. For example, a facial recognition algorithm may perform poorly when image resolution is low or images are taken in low lighting. Or a speech-to-text system might not be used reliably to provide closed captions for online lectures because it fails to handle technical jargon.
        \item The authors should discuss the computational efficiency of the proposed algorithms and how they scale with dataset size.
        \item If applicable, the authors should discuss possible limitations of their approach to address problems of privacy and fairness.
        \item While the authors might fear that complete honesty about limitations might be used by reviewers as grounds for rejection, a worse outcome might be that reviewers discover limitations that aren't acknowledged in the paper. The authors should use their best judgment and recognize that individual actions in favor of transparency play an important role in developing norms that preserve the integrity of the community. Reviewers will be specifically instructed to not penalize honesty concerning limitations.
    \end{itemize}

\item {\bf Theory assumptions and proofs}
    \item[] Question: For each theoretical result, does the paper provide the full set of assumptions and a complete (and correct) proof?
    \item[] Answer: \answerYes{} 
    \item[] Justification: it does 
    \item[] Guidelines:
    \begin{itemize}
        \item The answer \answerNA{} means that the paper does not include theoretical results. 
        \item All the theorems, formulas, and proofs in the paper should be numbered and cross-referenced.
        \item All assumptions should be clearly stated or referenced in the statement of any theorems.
        \item The proofs can either appear in the main paper or the supplemental material, but if they appear in the supplemental material, the authors are encouraged to provide a short proof sketch to provide intuition. 
        \item Inversely, any informal proof provided in the core of the paper should be complemented by formal proofs provided in appendix or supplemental material.
        \item Theorems and Lemmas that the proof relies upon should be properly referenced. 
    \end{itemize}

    \item {\bf Experimental result reproducibility}
    \item[] Question: Does the paper fully disclose all the information needed to reproduce the main experimental results of the paper to the extent that it affects the main claims and/or conclusions of the paper (regardless of whether the code and data are provided or not)?
    \item[] Answer: \answerYes{} 
    \item[] Justification: it does
    \item[] Guidelines:
    \begin{itemize}
        \item The answer \answerNA{} means that the paper does not include experiments.
        \item If the paper includes experiments, a \answerNo{} answer to this question will not be perceived well by the reviewers: Making the paper reproducible is important, regardless of whether the code and data are provided or not.
        \item If the contribution is a dataset and\slash or model, the authors should describe the steps taken to make their results reproducible or verifiable. 
        \item Depending on the contribution, reproducibility can be accomplished in various ways. For example, if the contribution is a novel architecture, describing the architecture fully might suffice, or if the contribution is a specific model and empirical evaluation, it may be necessary to either make it possible for others to replicate the model with the same dataset, or provide access to the model. In general. releasing code and data is often one good way to accomplish this, but reproducibility can also be provided via detailed instructions for how to replicate the results, access to a hosted model (e.g., in the case of a large language model), releasing of a model checkpoint, or other means that are appropriate to the research performed.
        \item While NeurIPS does not require releasing code, the conference does require all submissions to provide some reasonable avenue for reproducibility, which may depend on the nature of the contribution. For example
        \begin{enumerate}
            \item If the contribution is primarily a new algorithm, the paper should make it clear how to reproduce that algorithm.
            \item If the contribution is primarily a new model architecture, the paper should describe the architecture clearly and fully.
            \item If the contribution is a new model (e.g., a large language model), then there should either be a way to access this model for reproducing the results or a way to reproduce the model (e.g., with an open-source dataset or instructions for how to construct the dataset).
            \item We recognize that reproducibility may be tricky in some cases, in which case authors are welcome to describe the particular way they provide for reproducibility. In the case of closed-source models, it may be that access to the model is limited in some way (e.g., to registered users), but it should be possible for other researchers to have some path to reproducing or verifying the results.
        \end{enumerate}
    \end{itemize}

\item {\bf Open access to data and code}
    \item[] Question: Does the paper provide open access to the data and code, with sufficient instructions to faithfully reproduce the main experimental results, as described in supplemental material?
    \item[] Answer: \answerYes{} 
    \item[] Justification: it does
    \item[] Guidelines:
    \begin{itemize}
        \item The answer \answerNA{} means that paper does not include experiments requiring code.
        \item Please see the NeurIPS code and data submission guidelines (\url{https://neurips.cc/public/guides/CodeSubmissionPolicy}) for more details.
        \item While we encourage the release of code and data, we understand that this might not be possible, so \answerNo{} is an acceptable answer. Papers cannot be rejected simply for not including code, unless this is central to the contribution (e.g., for a new open-source benchmark).
        \item The instructions should contain the exact command and environment needed to run to reproduce the results. See the NeurIPS code and data submission guidelines (\url{https://neurips.cc/public/guides/CodeSubmissionPolicy}) for more details.
        \item The authors should provide instructions on data access and preparation, including how to access the raw data, preprocessed data, intermediate data, and generated data, etc.
        \item The authors should provide scripts to reproduce all experimental results for the new proposed method and baselines. If only a subset of experiments are reproducible, they should state which ones are omitted from the script and why.
        \item At submission time, to preserve anonymity, the authors should release anonymized versions (if applicable).
        \item Providing as much information as possible in supplemental material (appended to the paper) is recommended, but including URLs to data and code is permitted.
    \end{itemize}

\item {\bf Experimental setting/details}
    \item[] Question: Does the paper specify all the training and test details (e.g., data splits, hyperparameters, how they were chosen, type of optimizer) necessary to understand the results?
    \item[] Answer: \answerYes{} 
    \item[] Justification: it does
    \item[] Guidelines:
    \begin{itemize}
        \item The answer \answerNA{} means that the paper does not include experiments.
        \item The experimental setting should be presented in the core of the paper to a level of detail that is necessary to appreciate the results and make sense of them.
        \item The full details can be provided either with the code, in appendix, or as supplemental material.
    \end{itemize}

\item {\bf Experiment statistical significance}
    \item[] Question: Does the paper report error bars suitably and correctly defined or other appropriate information about the statistical significance of the experiments?
    \item[] Answer: \answerYes{} 
    \item[] Justification: it does
    \item[] Guidelines:
    \begin{itemize}
        \item The answer \answerNA{} means that the paper does not include experiments.
        \item The authors should answer \answerYes{} if the results are accompanied by error bars, confidence intervals, or statistical significance tests, at least for the experiments that support the main claims of the paper.
        \item The factors of variability that the error bars are capturing should be clearly stated (for example, train/test split, initialization, random drawing of some parameter, or overall run with given experimental conditions).
        \item The method for calculating the error bars should be explained (closed form formula, call to a library function, bootstrap, etc.)
        \item The assumptions made should be given (e.g., Normally distributed errors).
        \item It should be clear whether the error bar is the standard deviation or the standard error of the mean.
        \item It is OK to report 1-sigma error bars, but one should state it. The authors should preferably report a 2-sigma error bar than state that they have a 96\% CI, if the hypothesis of Normality of errors is not verified.
        \item For asymmetric distributions, the authors should be careful not to show in tables or figures symmetric error bars that would yield results that are out of range (e.g., negative error rates).
        \item If error bars are reported in tables or plots, the authors should explain in the text how they were calculated and reference the corresponding figures or tables in the text.
    \end{itemize}

\item {\bf Experiments compute resources}
    \item[] Question: For each experiment, does the paper provide sufficient information on the computer resources (type of compute workers, memory, time of execution) needed to reproduce the experiments?
    \item[] Answer: \answerYes{} 
    \item[] Justification: it does
    \item[] Guidelines:
    \begin{itemize}
        \item The answer \answerNA{} means that the paper does not include experiments.
        \item The paper should indicate the type of compute workers CPU or GPU, internal cluster, or cloud provider, including relevant memory and storage.
        \item The paper should provide the amount of compute required for each of the individual experimental runs as well as estimate the total compute. 
        \item The paper should disclose whether the full research project required more compute than the experiments reported in the paper (e.g., preliminary or failed experiments that didn't make it into the paper). 
    \end{itemize}
    
\item {\bf Code of ethics}
    \item[] Question: Does the research conducted in the paper conform, in every respect, with the NeurIPS Code of Ethics \url{https://neurips.cc/public/EthicsGuidelines}?
    \item[] Answer: \answerYes{} 
    \item[] Justification: we reviewed it 
    \item[] Guidelines:
    \begin{itemize}
        \item The answer \answerNA{} means that the authors have not reviewed the NeurIPS Code of Ethics.
        \item If the authors answer \answerNo, they should explain the special circumstances that require a deviation from the Code of Ethics.
        \item The authors should make sure to preserve anonymity (e.g., if there is a special consideration due to laws or regulations in their jurisdiction).
    \end{itemize}

\item {\bf Broader impacts}
    \item[] Question: Does the paper discuss both potential positive societal impacts and negative societal impacts of the work performed?
    \item[] Answer: \answerYes{} 
    \item[] Justification: help practitioners and theoreticians to better train RL model on repeated auctions with dynamic value. 
    \item[] Guidelines:
    \begin{itemize}
        \item The answer \answerNA{} means that there is no societal impact of the work performed.
        \item If the authors answer \answerNA{} or \answerNo, they should explain why their work has no societal impact or why the paper does not address societal impact.
        \item Examples of negative societal impacts include potential malicious or unintended uses (e.g., disinformation, generating fake profiles, surveillance), fairness considerations (e.g., deployment of technologies that could make decisions that unfairly impact specific groups), privacy considerations, and security considerations.
        \item The conference expects that many papers will be foundational research and not tied to particular applications, let alone deployments. However, if there is a direct path to any negative applications, the authors should point it out. For example, it is legitimate to point out that an improvement in the quality of generative models could be used to generate Deepfakes for disinformation. On the other hand, it is not needed to point out that a generic algorithm for optimizing neural networks could enable people to train models that generate Deepfakes faster.
        \item The authors should consider possible harms that could arise when the technology is being used as intended and functioning correctly, harms that could arise when the technology is being used as intended but gives incorrect results, and harms following from (intentional or unintentional) misuse of the technology.
        \item If there are negative societal impacts, the authors could also discuss possible mitigation strategies (e.g., gated release of models, providing defenses in addition to attacks, mechanisms for monitoring misuse, mechanisms to monitor how a system learns from feedback over time, improving the efficiency and accessibility of ML).
    \end{itemize}
    
\item {\bf Safeguards}
    \item[] Question: Does the paper describe safeguards that have been put in place for responsible release of data or models that have a high risk for misuse (e.g., pre-trained language models, image generators, or scraped datasets)?
    \item[] Answer: \answerNA{} 
    \item[] Justification: no release of pre trained LLM or data. 
    \item[] Guidelines:
    \begin{itemize}
        \item The answer \answerNA{} means that the paper poses no such risks.
        \item Released models that have a high risk for misuse or dual-use should be released with necessary safeguards to allow for controlled use of the model, for example by requiring that users adhere to usage guidelines or restrictions to access the model or implementing safety filters. 
        \item Datasets that have been scraped from the Internet could pose safety risks. The authors should describe how they avoided releasing unsafe images.
        \item We recognize that providing effective safeguards is challenging, and many papers do not require this, but we encourage authors to take this into account and make a best faith effort.
    \end{itemize}

\item {\bf Licenses for existing assets}
    \item[] Question: Are the creators or original owners of assets (e.g., code, data, models), used in the paper, properly credited and are the license and terms of use explicitly mentioned and properly respected?
    \item[] Answer: \answerYes{} 
    \item[] Justification: we are the creator of this simulator
    \item[] Guidelines:
    \begin{itemize}
        \item The answer \answerNA{} means that the paper does not use existing assets.
        \item The authors should cite the original paper that produced the code package or dataset.
        \item The authors should state which version of the asset is used and, if possible, include a URL.
        \item The name of the license (e.g., CC-BY 4.0) should be included for each asset.
        \item For scraped data from a particular source (e.g., website), the copyright and terms of service of that source should be provided.
        \item If assets are released, the license, copyright information, and terms of use in the package should be provided. For popular datasets, \url{paperswithcode.com/datasets} has curated licenses for some datasets. Their licensing guide can help determine the license of a dataset.
        \item For existing datasets that are re-packaged, both the original license and the license of the derived asset (if it has changed) should be provided.
        \item If this information is not available online, the authors are encouraged to reach out to the asset's creators.
    \end{itemize}

\item {\bf New assets}
    \item[] Question: Are new assets introduced in the paper well documented and is the documentation provided alongside the assets?
    \item[] Answer: \answerYes{} 
    \item[] Justification: it is the case
    \item[] Guidelines:
    \begin{itemize}
        \item The answer \answerNA{} means that the paper does not release new assets.
        \item Researchers should communicate the details of the dataset\slash code\slash model as part of their submissions via structured templates. This includes details about training, license, limitations, etc. 
        \item The paper should discuss whether and how consent was obtained from people whose asset is used.
        \item At submission time, remember to anonymize your assets (if applicable). You can either create an anonymized URL or include an anonymized zip file.
    \end{itemize}

\item {\bf Crowdsourcing and research with human subjects}
    \item[] Question: For crowdsourcing experiments and research with human subjects, does the paper include the full text of instructions given to participants and screenshots, if applicable, as well as details about compensation (if any)? 
    \item[] Answer: \answerNA{} 
    \item[] Justification: 
    \item[] Guidelines:
    \begin{itemize}
        \item The answer \answerNA{} means that the paper does not involve crowdsourcing nor research with human subjects.
        \item Including this information in the supplemental material is fine, but if the main contribution of the paper involves human subjects, then as much detail as possible should be included in the main paper. 
        \item According to the NeurIPS Code of Ethics, workers involved in data collection, curation, or other labor should be paid at least the minimum wage in the country of the data collector. 
    \end{itemize}

\item {\bf Institutional review board (IRB) approvals or equivalent for research with human subjects}
    \item[] Question: Does the paper describe potential risks incurred by study participants, whether such risks were disclosed to the subjects, and whether Institutional Review Board (IRB) approvals (or an equivalent approval/review based on the requirements of your country or institution) were obtained?
    \item[] Answer: \answerNA{} 
    \item[] Justification: 
    \item[] Guidelines:
    \begin{itemize}
        \item The answer \answerNA{} means that the paper does not involve crowdsourcing nor research with human subjects.
        \item Depending on the country in which research is conducted, IRB approval (or equivalent) may be required for any human subjects research. If you obtained IRB approval, you should clearly state this in the paper. 
        \item We recognize that the procedures for this may vary significantly between institutions and locations, and we expect authors to adhere to the NeurIPS Code of Ethics and the guidelines for their institution. 
        \item For initial submissions, do not include any information that would break anonymity (if applicable), such as the institution conducting the review.
    \end{itemize}

\item {\bf Declaration of LLM usage}
    \item[] Question: Does the paper describe the usage of LLMs if it is an important, original, or non-standard component of the core methods in this research? Note that if the LLM is used only for writing, editing, or formatting purposes and does \emph{not} impact the core methodology, scientific rigor, or originality of the research, declaration is not required.
    \item[] Answer: \answerNA{} 
    \item[] Justification: only used for writing and code cleaning purpose. 
    \item[] Guidelines:
    \begin{itemize}
        \item The answer \answerNA{} means that the core method development in this research does not involve LLMs as any important, original, or non-standard components.
        \item Please refer to our LLM policy in the NeurIPS handbook for what should or should not be described.
    \end{itemize}

\end{enumerate}

\end{document}